\documentclass[
  reprint,            
  aps,prb,            
  superscriptaddress, 
  longbibliography
]{revtex4-2}

\usepackage{graphicx}                 
\usepackage{amsmath,amssymb,bm}       
\renewcommand\Re{\operatorname{Re}}

\renewcommand\P{\operatorname{P}}
\usepackage{siunitx}                  
\usepackage{xcolor}
\usepackage[utf8]{inputenc}
\usepackage{mathtools}
\usepackage{physics}
\DeclareSIUnit\ryd{Ry}
\DeclareSIUnit\fs{fs}
\usepackage[version=4]{mhchem}
\usepackage{booktabs}
\usepackage[T1]{fontenc}

\usepackage[colorlinks=true,allcolors=blue]{hyperref}

\graphicspath{{figs/}}

\begin{document}

\title{Coherent and Incoherent Time-Resolved Diffuse Scattering Intensity from First Principles}

\author{Matteo Furci}
\email{matteo.furci@unitn.it}
\affiliation{Department of Physics, University of Trento, Via Sommarive 14, 38123 Povo, Italy}
\author{Giovanni Marini}
\affiliation{Department of Physics, University of Trento, Via Sommarive 14, 38123 Povo, Italy}
\author{Stefano Mocatti}
\affiliation{Department of Physics, University of Trento, Via Sommarive 14, 38123 Povo, Italy}
\author{Matteo Calandra}
\email{m.calandrabuonaura@unitn.it}
\affiliation{Department of Physics, University of Trento, Via Sommarive 14, 38123 Povo, Italy}
\date{\today}

\begin{abstract}
Time-resolved diffuse scattering from X-ray free-electron lasers and ultrafast electron diffraction gives access to the nonequilibrium lattice dynamics of photoexcited solids in reciprocal space. Existing interpretations have often relied on simplified models. Here we present a first-principles formulation of the time-dependent first-order diffuse intensity in which coherent and incoherent contributions emerge on equal footing. By assuming the probe-pulse coherence time to be longer than the characteristic electronic timescales, but shorter than those of the ionic motion, we work in the density-matrix formalism and derive equations of motion for the displacement-displacement correlators in a basis of instantaneous, out-of-equilibrium Born-Oppenheimer normal modes. We show how the incoherent contribution, governed by the phonon occupations, and a coherent contribution, governed by the anomalous phonon correlators, emerge. The non-trivial coherent dynamics is shown to be induced both by the photoinduced phonon-frequency shift and by electron-phonon scattering off the nonequilibrium carrier distribution. We implement the resulting equations using Wannier interpolation in the \textsc{EPI}q code and apply them to photoexcited germanium. The calculated diffuse intensity reproduces both the oscillation frequencies and the rising incoherent background observed experimentally, opening the way to a systematic interpretation of time-resolved diffuse scattering in ultrafast experiments.
\end{abstract}

\maketitle

\section{Introduction}
 
The investigation of lattice dynamics is fundamental to understanding the macroscopic properties of solids, from thermal conductivity to structural phase transitions. 
X-ray and neutron scattering are leading experimental techniques for measuring phonon dispersion relations, as they provide access to structural information via the Bragg peaks and to vibrational properties via the thermal diffuse background. 
Because the thermal diffuse scattering cross section is far smaller than that of the Bragg peaks, this connection became broadly exploitable with the advent of bright X-ray sources.
At equilibrium, and within the Born-Oppenheimer and harmonic approximations, the first-order thermal diffuse scattering intensity is proportional to the atomic displacement-displacement correlation function, so that fitting force-constant models to the measured pattern yields the eigenfrequencies and eigenvectors characterizing the vibrational spectrum~\cite{Laval,Born_1942,James_Book,Wu1999,Zacharias2021}. 
 
Recent advances have greatly expanded our ability to perform femtosecond-resolved experiments. 
High-brightness X-ray free-electron lasers (XFELs) and ultrafast electron diffraction (UED) sources can be used in a pump-probe configuration to monitor the thermal diffuse scattering intensity across large volumes of reciprocal space with sub-picosecond temporal resolution~\cite{Trigo2013}. 
Ultrashort probe pulses reduce temporal averaging enough to capture nearly instantaneous snapshots of the displacement-displacement correlators, opening the possibility of directly probing the phonon spectrum of a photoexcited, out-of-equilibrium system. 
A characteristic dynamical signature is an oscillation of the thermal diffuse scattering intensity at a given momentum transfer at twice the phonon frequency of the corresponding reduced wavevector. The Fourier transform of the time trace therefore yields the phonon dispersion, with an energy resolution primarily set by the length of the delay scan~\cite{Zhu2015}.
 
Time-resolved diffuse signals contain a great deal of information about ultrafast atomic motion, and it is now well established that they exhibit two qualitatively distinct types of temporal modulation. 
The first is comparatively slow, developing over a few picoseconds, and reflects the gradual heating of the lattice as electrons and phonons exchange energy following the pump: the phonon occupations increase and the diffuse intensity grows with the amplitude of the thermal fluctuations~\cite{Stern_2018,deCotret_2019}. 
The second is a faster oscillatory component superimposed on that background, which carries direct information on specific phonon modes and signals the emergence of coherent phonon correlations. 
Both are clearly visible, as exemplified by the measurements on germanium samples of Refs.~\onlinecite{Trigo2013} and~\onlinecite{Zhu2015}, where the Fourier transform of the oscillatory component shows dominant weight at twice the frequency of the lowest phonon branches.
 
Subsequent experiments established time-resolved diffuse scattering as a probe of the coupling between the lattice and other degrees of freedom and revealed features that go beyond a simple photoinduced frequency renormalization. 
In PbTe, additional spectral components appear that cannot be assigned to the double-frequency overtone of a single branch, but instead correspond to sum and difference frequencies of transverse optical and transverse acoustic modes~\cite{Jiang2016}. 
These features were interpreted as evidence of the photoinduced modification and coupling of phonon eigenmodes. 
In bismuth, the usual overtone signal coexists with additional weight on branches satisfying the resonance condition $2\omega_{\mathbf{q}\nu} = \omega_{A_{1g}}$ with the coherently driven $A_{1g}$ optical mode, a signature of strong anharmonic decay of that mode into correlated phonon pairs at $\mathbf{q}$ and $-\mathbf{q}$~\cite{Teitelbaum2018}. 
Related measurements have resolved phonon squeezing, anisotropic lattice cooling, and the decay of coherent optical phonons into acoustic modes~\cite{PhysRevB.101.100302}, uncovered the real-time dynamics of the ultrafast structural phase transition in VO$_2$~\cite{doi:10.1126/science.aau3873}, and identified a nonthermal bonding origin~\cite{Huang2023} for the photoexcited lattice instability of SnSe~\cite{Huang2022,Mocatti2023,Dringoli2024}.
 
The theoretical description of these experiments has often treated the coherent and incoherent components of the signal using separate approaches.
The incoherent background is usually described by the two-temperature model of Allen~\cite{Allen1987} and its multi-temperature extensions. In this approach, the lattice is assigned one or a few effective temperatures whose evolution follows rate equations with an electron-phonon coupling constant expressible through the Eliashberg function~\cite{Caruso31122022}, or by more refined Boltzmann equation approaches tracking the dynamical evolution of electronic and phononic occupations in time~\cite{Britt2022-ut,Murphy-Armando2023-ld}. More recently, first-principles nonequilibrium formulations have been developed that describe phonon occupations and anomalous phonon correlations on the same footing~\cite{Stefanucci2024}.
Population-based approaches give an intuitive and often quantitatively adequate account of the slow rise of the diffuse intensity, but do not by themselves describe the coherent oscillations.
A widely used description of the oscillatory component is the sudden-quench model of Johnson \emph{et al.}~\cite{Johnson2009}, building on the treatment of harmonic oscillators with time-dependent parameters of Kiss \emph{et al.}~\cite{Kiss1994} and on the identification of light-driven squeezed lattice states in second-order Raman scattering~\cite{Garrett1997}.
Within these frameworks, the normal-mode frequencies are assumed to change instantaneously from their ground-state to their photoexcited values, and the resulting impulsive squeezing of the vibrational wavepacket produces an oscillation of its variance, and hence of the diffuse intensity, at twice the new frequency, while the mean displacement is unchanged.
This model picture, while qualitative, gives a useful description of the coherent oscillations and has served as a standard interpretive guide for experiments~\cite{Trigo_2018}.
 
To enable a systematic interpretation of these measurements, in this work we present a first-principles framework for the time-resolved diffuse signal that treats both components on the same footing.
Working within density-matrix theory and the electron-phonon Bloch equations~\cite{Rossi2002}, we obtain equations of motion for the phonon correlators and discuss the physical origin of both the coherent and incoherent components of the signal, recovering the sudden-quench picture as a limiting case. 
We implement the scheme in the \textsc{EPI}q code~\cite{Marini2024} and validate it against the femtosecond diffuse scattering measurements on germanium of Ref.~\onlinecite{Zhu2015}.
 
The paper is organized as follows. Section~\ref{sec:cross-section} derives the small-displacement expansion of the time-dependent scattering cross section and identifies the term responsible for the diffuse signal.
Section~\ref{sec:eom} derives the equations of motion for the target correlators and specializes them to the no-mode-mixing case.
Section~\ref{sec:implementation} describes the implementation, and Sec.~\ref{sec:results} presents the comparison with experiment. 
The sudden-quench limit and its extension to mode mixing are discussed in Appendix~\ref{app:sudden}, while Appendix~\ref{app:bragg} discusses the contribution of coherent $\Gamma$-point phonons. The general mode-mixing equations and the closure and Markov approximations underlying Sec.~\ref{sec:eom} are reported in Appendices~\ref{app:mixing-eqs}, \ref{app:closure}, and~\ref{app:markov}, respectively.

\section{Time-Dependent Scattering Cross Section}
\label{sec:cross-section}

The time-dependent X-ray or electron scattering intensity can be obtained from standard first-order time-dependent perturbation theory. 
If the coherence time of the probe pulse is much longer than the characteristic electronic dynamics timescales, but still much shorter than those of the ionic motion, it can be shown~\cite{Henriksen2008,Furci2026} that the scattering differential cross section takes the form
\begin{equation}
    \frac{d\sigma}{d\Omega_\text{s} d\omega_{\mathbf{k}_\text{s}}}
    \propto \text{tr}\left\{ \hat{L}^\dagger (\mathbf{Q}) \hat{L} (\mathbf{Q}) \hat{\rho}_\text{N}(t) \right\},
\label{eq:cross-section}
\end{equation}
where $\hat{L}^\dagger (\mathbf{Q}) \hat{L} (\mathbf{Q})$ is the scattering operator, with $\mathbf{Q}$ denoting the momentum exchanged in the scattering event. In Eq.~\eqref{eq:cross-section}, $\hat{\rho}_\text{N}(t)$ denotes the time-dependent density operator for the ionic degrees of freedom.
For a crystalline sample, within the independent-atom approximation~\cite{James_Book},
\begin{equation}
    \hat{L}(\mathbf{Q}) = \sum_{Aa} \mathcal{F}_a(\mathbf{Q}) \exp\left\{i \mathbf{Q} \cdot (\mathbf{R}_A  + \bm{\tau}_a + \mathbf{u}_{Aa}) \right\},
\label{eq:L-operator}
\end{equation}
where $\mathcal{F}_a(\mathbf{Q})$ is the probe-dependent atomic form factor, $\mathbf{R}_A +
\bm{\tau}_a$ is the equilibrium coordinate of atom $a$ in the $A^{\rm th}$ unit cell, and
$\mathbf{u}_{Aa}$ is its displacement from equilibrium. For notational simplicity, we define in the following $f_a(\mathbf{Q})=e^{i\mathbf{Q}\cdot \bm{\tau}_a}\mathcal{F}_a(\mathbf{Q})$, absorbing atomic cell phase factors $e^{i\mathbf{Q}\cdot \bm{\tau}_a}$ into the atomic structure factor.
Under the assumption of small ionic displacements, the dominant contributions
follow from a small-displacement expansion,
\begin{widetext}
\begin{align}
\nonumber    &\langle\hat{L}^\dagger (\mathbf{Q}) \hat{L} (\mathbf{Q})\rangle_t =  \left| \sum_{Aa} f_a(\mathbf{Q})e^{i \mathbf{Q} \cdot \mathbf{R}_A} \right|^2 \, + \,  2N\sqrt{N}\delta_{\mathbf{Q}, \mathbf{G}} \sum_\nu \Re\left\{ i\langle u_{\Gamma \nu}\rangle_t \sum_{aa'} f_{a}(\mathbf{G}) f^*_{a'}(\mathbf{G}) \, (\mathbf{G}\cdot \tilde{\mathbf{e}}^a_{\Gamma\nu})  \right\} \,+ \\
\nonumber &\hspace{73pt}-N \delta_{\mathbf{Q}, \mathbf{G}} \sum_{aa'\nu\nu'} f_{a}(\mathbf{G}) f^*_{a'}(\mathbf{G})  \sum_{\mathbf{q}}\langle u_{\mathbf{q}\nu} u_{-\mathbf{q}\nu'} \rangle_t (\mathbf{G}\cdot \tilde{\mathbf{e}}^a_{\mathbf{q}\nu})(\mathbf{G}\cdot \tilde{\mathbf{e}}^{a'}_{-\mathbf{q}\nu'}) \, + \\
&\hspace{73pt}+N \sum_{aa'\nu\nu'}  f_{a}(\mathbf{Q}) f^*_{a'}(\mathbf{Q}) \langle u_{\mathbf{Q}\nu} u_{-\mathbf{Q}\nu'} \rangle_t (\mathbf{Q}\cdot \tilde{\mathbf{e}}^{a}_{\mathbf{Q}\nu})(\mathbf{Q}\cdot \tilde{\mathbf{e}}^{a'}_{-\mathbf{Q}\nu'}) \, + \, \mathcal{O}{\left\{u^3_{Aa\alpha}\right\}},
\label{eq:expansion}
\end{align}
\end{widetext}
where $\tilde{\mathbf{e}}^{a}_{\mathbf{Q}\nu}$ is the mass-rescaled projection of the normal-mode eigenvector onto the $a$th atom in the unit cell, defined by $\tilde{\mathbf{e}}^{\,a}_{\mathbf{q}\nu} =\frac{\mathbf{e}^{\,a}_{\mathbf{q}\nu}}{\sqrt{M_{a}}}$. Here, $\mathbf{e}^{\,a}_{\mathbf{q}\nu}$ is the phonon eigenvector obtained by diagonalizing the Fourier transform of the dynamical matrix,
\begin{equation}
    \sum_{b\beta}D^{\alpha\beta}_{ab}(\mathbf{q})\,e^{b\beta}_{\mathbf{q}\nu}
      =\omega^{2}_{\mathbf{q}\nu}\,e^{a\alpha}_{\mathbf{q}\nu}    
\end{equation}
where the dynamical matrix is defined as
\begin{equation}
D^{\alpha\beta}_{ab}(\mathbf{q})=\frac{1}{\sqrt{M_{a}M_{b}}}
      \sum_{B}C^{\alpha\beta}_{0a,Bb}\,e^{i\mathbf{q}\cdot\mathbf{R}_{B}}
\end{equation}
and $C^{\alpha\beta}_{0a,Bb}$ is the force-constant matrix.

The quantity $u_{\mathbf{q}\nu}$ is the Fourier amplitude entering the normal-mode decomposition of the atomic displacements,
\begin{equation}
\mathbf{u}_{Aa} = \frac{1}{\sqrt{N}}\sum_{\mathbf{q}\nu}
      u_{\mathbf{q}\nu} \,\tilde{\mathbf{e}}^{\,a}_{\mathbf{q}\nu}\,
      e^{i\mathbf{q}\cdot\mathbf{R}_{A}}.
\end{equation}
Provided that photoexcitation does not break the translational symmetry of the crystal, the displacement-displacement correlators retain reciprocal-lattice periodicity. Hence, for $\mathbf{Q}=\mathbf{G}+\mathbf{q}$, they can be expressed in terms of the corresponding reduced wavevector $\mathbf{q}$.

In Eq.~\eqref{eq:expansion}, the subscript $t$ indicates that the expectation value is evaluated at the central time $t$ of the probe pulse. In this equation, one recognizes the time-dependent counterparts of the standard expressions for thermal diffuse scattering. 
The first term represents the standard Bragg intensity contribution; the second describes the effect of coherent phonon generation on the Bragg peaks and is nonzero only when $\mathbf{Q}$ is a reciprocal lattice vector and for modes with a nonvanishing $\Gamma$-point displacement expectation value~\cite{Zeiger1992} (see Appendix~\ref{app:bragg}); the third is the leading time-dependent Debye-Waller correction to the Bragg peaks, already discussed within a simplified model~\cite{Johnson2009}. 
Finally, the fourth term is the quantity of interest here: the time-dependent first-order diffuse intensity observed away from Bragg peaks in ultrafast diffraction experiments~\cite{Trigo2013,Zhu2015,doi:10.1126/science.aau3873}. 
The mode-mixing (off-diagonal) terms involving cross-correlations between different normal modes are retained at this level of the formulation.

Calculating the time-dependent scattering intensity from X-rays or electrons therefore requires the \emph{ab initio} determination of the coherent displacement amplitude $\langle u_{\Gamma \nu}\rangle_t$~\cite{Zeiger1992} and of the displacement-displacement correlators $\langle u_{\mathbf{Q}\nu}u_{-\mathbf{Q}\nu'} \rangle_t$. 
In this work, we neglect the contribution arising from the coherent phonon at $\Gamma$, $\langle u_{\Gamma \nu}\rangle_t$, leaving its analysis to future work (see also Appendix~\ref{app:bragg}).

\section{Equations of Motion for the Correlators}
\label{sec:eom}

\subsection{Electron-Phonon Hamiltonian in the Instantaneous Normal-Mode Basis}

A direct route to the time dependence of the target correlators is provided by the density-matrix formalism~\cite{Rossi2002}, which describes the dynamical evolution of the observables of interest via Heisenberg equations of motion and produces a closed set of equations once a truncation scheme for the resulting infinite hierarchy is chosen.

Since we are interested primarily in the vibrational degrees of freedom rather than in the electronic dynamics, we describe the system through the electron-phonon Hamiltonian~\cite{Giustino2017}, in which the phonon normal modes are taken to be those that instantaneously diagonalize the dynamical matrix associated with the time-dependent, out-of-equilibrium Born-Oppenheimer potential,
\begin{align}
\nonumber    \hat{H}(t) &= \sum_{n\mathbf{k}} \epsilon_{n\mathbf{k}} \hat{c}^\dagger_{n\mathbf{k}}\hat{c}_{n\mathbf{k}} + \sum_{\mathbf{q}\nu} \hbar\omega_{\mathbf{q}\nu} (t)\left(\hat{a}^\dagger_{\mathbf{q}\nu}(t)\hat{a}_{ \mathbf{q}\nu}(t) +\tfrac{1}{2}\right) + \\
&\hspace{-5pt}+ \sum_{mn\mathbf{k}\mathbf{q}\nu} g^\nu_{mn}(t, \mathbf{k}, \mathbf{q}) \hat{c}^\dagger_{m\mathbf{k}+\mathbf{q}}\hat{c}_{n\mathbf{k}}\left(\hat{a}_{ \mathbf{q}\nu}(t) + \hat{a}^\dagger_{ -\mathbf{q}\nu}(t)\right).
\label{eq:hamiltonian}
\end{align}
In the general formulation, the time dependence of $\omega_{\mathbf{q}\nu}(t)$ and $g^\nu_{mn}(t)$ is inherited from the evolving electronic occupations, which modify the screening of the interatomic force constants. The time dependence of the electronic eigenvalues is not explicitly considered, and we fix them to their photoexcited-state values. Equation ~\eqref{eq:hamiltonian} does not itself generate the electronic dynamics: the nonequilibrium carrier distribution $f_{n\mathbf{k}}(t)$ enters as an external input. This reflects the adiabatic separation underlying the Born-Oppenheimer treatment of photoexcited systems. A more detailed discussion of the relationship between this model Hamiltonian and the \emph{ab initio} one, as well as of the initial nonequilibrium electron distribution, can be found in Ref.~\cite{Mocatti2026}.

\subsection{Equation of Motion for the Displacement-Displacement Correlator}

From Eq.~\eqref{eq:expansion}, the target observables are the equal-time displacement-displacement correlators, which in the Heisenberg picture read
\begin{align}
\nonumber    &\langle u_{\mathbf{q}\nu} u_{-\mathbf{q}\nu'} \rangle_t = \frac{\hbar}{2\sqrt{\omega_{\mathbf{q}\nu}(t)\omega_{-\mathbf{q}\nu'}(t)}} \times \\
\nonumber& \hspace{10pt} \times \langle \hat{a}_{\mathbf{q}\nu}(t)\hat{a}_{-\mathbf{q}\nu'}(t) + \hat{a}_{\mathbf{q}\nu}(t)\hat{a}^\dagger_{\mathbf{q}\nu'}(t) \, + \\
&\hspace{23pt}+\,  \hat{a}^\dagger_{\mathbf{-q}\nu}(t)\hat{a}_{-\mathbf{q}\nu'}(t)   + \hat{a}^\dagger_{-\mathbf{q}\nu}(t)\hat{a}^\dagger_{\mathbf{q}\nu'}(t) \rangle .
\label{eq:target}
\end{align}
Up to an exchange of indices $\nu\leftrightarrow\nu'$, the last two terms follow from the first two by Hermitian conjugation, so it is sufficient to follow the dynamics of the latter. The first term, $\langle \hat{a}_{\mathbf{q}\nu}(t)\hat{a}_{-\mathbf{q}\nu'}(t)\rangle$, is an anomalous correlator: it vanishes in thermal equilibrium and carries the coherent oscillation of the diffuse intensity. The second term reduces to $n_{\mathbf{q}\nu}(t)+1$ in the absence of mode mixing and constitutes the incoherent background.

Applying the Heisenberg equations of motion and closing the hierarchy through a cluster expansion of the four-point correlators followed by a Markov approximation (Appendices~\ref{app:closure} and~\ref{app:markov}), we obtain the general result reported in Appendix~\ref{app:mixing-eqs}: Eq.~\eqref{eq:eom-mixing-1} for the first term and Eq.~\eqref{eq:eom-mixing-2} for the second. The evolution simplifies considerably if the phonon eigenvectors are assumed not to change significantly after photoexcitation (no-mode-mixing approximation). Under this assumption, the mixed-mode contributions can be neglected and only the $\nu'=\nu$ terms retained, yielding
\begin{widetext}
\begin{align}
\nonumber    \frac{d}{dt} \langle \hat{a}_{\mathbf{q}\nu}\hat{a}_{-\mathbf{q}\nu} \rangle =& \left[-i \left(\omega_{\mathbf{q}\nu}(t) + \omega_{-\mathbf{q}\nu}(t)\right) \right] \langle \hat{a}_{\mathbf{q}\nu}\hat{a}_{-\mathbf{q}\nu} \rangle
+ \frac{1}{2} \frac{\dot{\omega}_{\mathbf{q}\nu}(t)}{\omega_{\mathbf{q}\nu}(t)}  n_{\mathbf{q}\nu}(t)   \, + \,  \frac{1}{2} \frac{\dot{\omega}_{-\mathbf{q}\nu}(t)}{\omega_{-\mathbf{q}\nu}(t)}  \left( n_{-\mathbf{q}\nu}(t) + 1 \right) \, + \, \\
\nonumber& \hspace{-40pt} +\frac{\pi}{\hbar} \sum_{mn\mathbf{k}} \left| g^\nu_{mn}(t, \mathbf{k}, \mathbf{q})\right|^2 \Bigg\{\left[ \left(f_{m\mathbf{k}+\mathbf{q}} - f_{n\mathbf{k}} \right) \left( \langle \hat{a}_{\mathbf{q}\nu} \hat{a}_{-\mathbf{q}\nu}  \rangle - n_{\mathbf{q}\nu}(t) \right) +   f_{m \mathbf{k}+\mathbf{q}} \left( f_{n\mathbf{k}} - 1 \right) \right]\delta(\epsilon_{m\mathbf{k}+\mathbf{q}} - \epsilon_{n\mathbf{k}} - \hbar\omega_{\mathbf{q}\nu}(t)) + \\
&  \hspace{30pt}+ \left[ \left(f_{m\mathbf{k}+\mathbf{q}}  -  f_{n\mathbf{k}}  \right) \left( n_{-\mathbf{q}\nu}(t)  - \langle \hat{a}_{\mathbf{q}\nu} \hat{a}_{-\mathbf{q}\nu}  \rangle \right) + f_{n\mathbf{k}} \left( f_{m \mathbf{k}+\mathbf{q}} - 1\right)\right]\delta(\epsilon_{m\mathbf{k}+\mathbf{q}} - \epsilon_{n\mathbf{k}} + \hbar\omega_{-\mathbf{q}\nu}(t))\Bigg\}
\label{eq:eom-aa}
\end{align}
\end{widetext}
and an analogous equation for $\langle \hat{a}_{\mathbf{q}\nu}
\hat{a}^\dagger_{\mathbf{q}\nu}\rangle$:
\begin{widetext}
\small{
\begin{align}
\nonumber    &\frac{d}{dt}\langle\hat{a}_{\mathbf{q}\nu}\hat{a}^\dagger_{\mathbf{q}\nu} \rangle = \frac{1}{2} \frac{\dot{\omega}_{\mathbf{q}\nu}(t)}{\omega_{\mathbf{q}\nu}(t)} \langle \hat{a}^\dagger_{\mathbf{q}\nu}\hat{a}^\dagger_{-\mathbf{q}\nu} \rangle \, + \, \frac{1}{2}\frac{\dot{\omega}_{-\mathbf{q}\nu}(t)}{\omega_{-\mathbf{q}\nu}(t)} \langle \hat{a}_{\mathbf{q}\nu}\hat{a}_{-\mathbf{q}\nu} \rangle \, + \,\\
\nonumber& +\frac{2\pi}{\hbar} \sum_{mn\mathbf{k}} \left|g^\nu_{mn}(t, \mathbf{k}, \mathbf{q})\right|^2 \Bigg\{\left[ \left(f_{m\mathbf{k}+\mathbf{q}} - f_{n\mathbf{k}} \right) n_{\mathbf{q}\nu}(t)  + f_{m \mathbf{k}+\mathbf{q}} \left(1 - f_{n\mathbf{k}} \right) \right]\delta(\epsilon_{m\mathbf{k}+\mathbf{q}} - \epsilon_{n\mathbf{k}} - \hbar\omega_{\mathbf{q}\nu}(t)) + \\
&  \hspace{110pt}+   \left(f_{m\mathbf{k}+\mathbf{q}} - f_{n\mathbf{k}} \right) \Re\left\{\langle \hat{a}_{\mathbf{q}\nu} \hat{a}_{-\mathbf{q}\nu}  \rangle\right\}\delta(\epsilon_{m\mathbf{k}+\mathbf{q}} - \epsilon_{n\mathbf{k}} + \hbar\omega_{-\mathbf{q}\nu}(t))\Bigg\}.
\label{eq:eom-adag}
\end{align}
}
\end{widetext}
The complete equations without the no-mode-mixing approximation are reported in Appendix~\ref{app:mixing-eqs}.

Equations~\eqref{eq:eom-aa} and~\eqref{eq:eom-adag} are the central result of this section. Assuming that time-reversal symmetry is preserved during the photoexcited dynamics, such that $\omega_{\mathbf{q}\nu}(t)=\omega_{-\mathbf{q}\nu}(t)$ and $n_{\mathbf{q}\nu}(t)=n_{-\mathbf{q}\nu}(t)$, Eq.~\eqref{eq:eom-aa} can be written in the compact form
\begin{align}
\nonumber \frac{d}{dt}\langle \hat{a}_{\mathbf{q}\nu} \hat{a}_{-\mathbf{q}\nu} \rangle
&= -2i\omega_{\mathbf{q}\nu}(t) \langle \hat{a}_{\mathbf{q}\nu} \hat{a}_{-\mathbf{q}\nu} \rangle \, + \\
&\hspace{10pt} - \Gamma_{\mathbf{q}\nu}(t) \langle \hat{a}_{\mathbf{q}\nu} \hat{a}_{-\mathbf{q}\nu} \rangle + F_{\mathbf{q}\nu}(t),
\label{eq:structure}
\end{align}
where the anomalous correlator precesses at twice the instantaneous phonon frequency, giving rise to the $2\omega_{\mathbf{q}\nu}$ periodicity of the measured intensity, while $\Gamma_{\mathbf{q}\nu}(t)$ is the electron-phonon damping rate. 
The source term contains two distinct contributions. 
The first, $\tfrac{1}{2}(\dot{\omega}_{\mathbf{q}\nu}/\omega_{\mathbf{q}\nu})(2n_{\mathbf{q}\nu}+1)$, is the squeezing drive: it is nonzero only while the mode frequency is changing and is thermally enhanced through the $(2n+1)$ factor. 
In the limit of an instantaneous frequency step, it integrates to the sudden-quench result of Appendix~\ref{app:sudden}, recovering the semi-empirical model of Ref.~\onlinecite{Johnson2009} as a limiting case. 
The second, associated with the terms $f_{m\mathbf{k}+\mathbf{q}}(f_{n\mathbf{k}}-1)$ and $f_{n\mathbf{k}}(f_{m\mathbf{k}+\mathbf{q}}-1)$, is an electron-phonon source term that survives even at constant frequency: a nonequilibrium carrier distribution can therefore generate anomalous phonon correlations directly, without a lattice instability. 
The relative weight of these two contributions is material- and mode-dependent and can be evaluated explicitly within the present framework.

We note that, for time-independent phonon frequencies, Eq.~\eqref{eq:eom-aa} reduces to the corresponding nonequilibrium Green-function result of Ref.~\onlinecite{Stefanucci2024}, while Eqs.~\eqref{eq:eom-mixing-1}-\eqref{eq:eom-mixing-2} provide its natural generalization to the mode-mixing case.

\subsection{Diffuse intensity}

Inserting Eq.~\eqref{eq:target} into the fourth term of
Eq.~\eqref{eq:expansion} and specializing to the no-mode-mixing case, the
first-order diffuse intensity at momentum transfer $\mathbf{Q} =
\mathbf{G}+\mathbf{q}$ becomes
\begin{equation}
	\begin{gathered}
I_1(\mathbf{Q}, t) \propto \sum_{\nu} \frac{\left| \sum_a f_a(\mathbf{Q})  \left(\mathbf{Q}\cdot\tilde{\mathbf{e}}^a_{\mathbf{q}\nu}\right)\right|^2}{2\omega_{\mathbf{q}\nu}(t)} \times\\
\left[ 2n_{\mathbf{q}\nu}(t) + 1 + 2\,\Re \langle \hat{a}_{\mathbf{q}\nu}\hat{a}_{-\mathbf{q}\nu}\rangle_t \right].
\label{eq:intensity}
	\end{gathered}
\end{equation}
The terms in square brackets separate the signal into two contributions: $2n_{\mathbf{q}\nu}(t)+1$ is the
incoherent contribution, which changes as energy flows from
the carriers into the lattice, while $2\Re\langle \hat{a}\hat{a}\rangle_t$
carries an oscillatory component with characteristic frequency $2\omega_{\mathbf{q}\nu}$. Equation~\eqref{eq:intensity} is evaluated at every time step of the implementation described next.

\begin{figure}[t]
\centering
\includegraphics[width=\columnwidth]{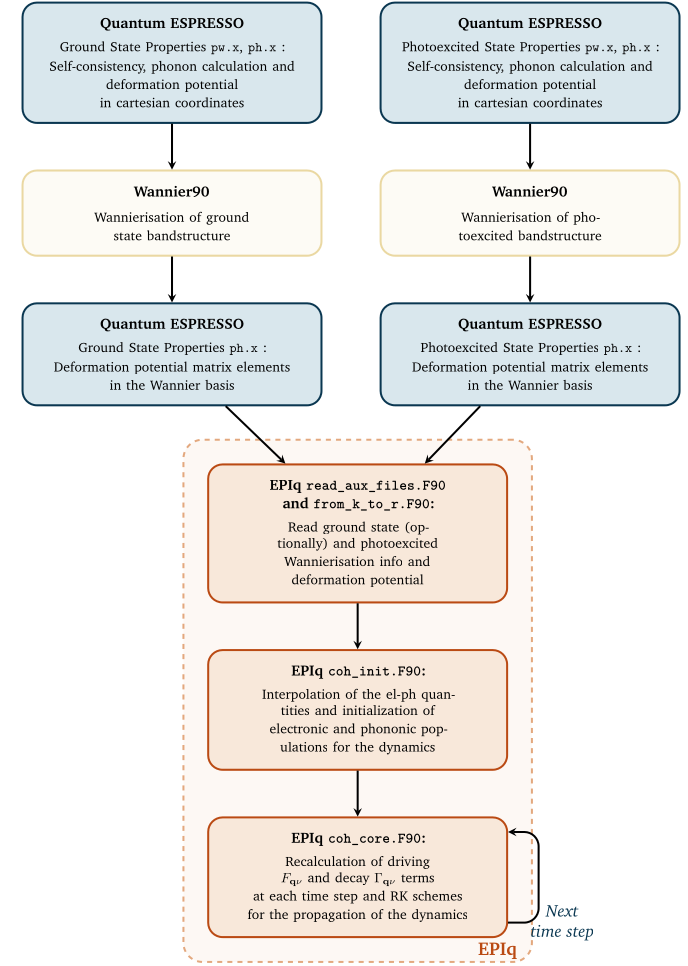}
\caption{Workflow of the time-dependent diffuse scattering implementation.
Ground-state and photoexcited electronic structures, phonons, and deformation
potentials are computed independently with Quantum ESPRESSO
(\texttt{pw.x}, \texttt{ph.x}) and Wannierized with Wannier90; the
deformation-potential matrix elements are then re-expressed in the Wannier
basis. Within \textsc{EPI}q, \texttt{read\_aux\_files.F90} and
\texttt{from\_k\_to\_r.F90} read both sets of quantities,
\texttt{coh\_init.F90} interpolates the electron-phonon matrix elements onto
the dynamics grids and initializes the electronic and phononic populations,
and \texttt{coh\_core.F90} recomputes the driving $F_{\mathbf{q}\nu}$ and
damping $\Gamma_{\mathbf{q}\nu}$ terms of Eq.~\eqref{eq:structure} at each time
step and propagates the dynamics using Runge--Kutta schemes. The shaded
box outlined by a dashed line indicates the modules introduced in the present work. The dynamical
propagation, including the RK integrator, was adapted from the implementation
of photoexcited carrier dynamics in \textsc{EPI}q~\cite{Mocatti2026}.}
\label{fig:workflow}
\end{figure}

\section{Implementation}
\label{sec:implementation}

The scheme has been implemented in the \textsc{EPI}q code~\cite{Marini2024}. The workflow is summarized in Fig.~\ref{fig:workflow}. Two separate first-principles calculations are required, one for the ground state and one for the photoexcited state. This is the main difference from a standard \textsc{EPI}q workflow, in which the Wannier interpolation and electron-phonon calculations are performed only once. 
The photoexcited calculation is carried out at fixed excited-carrier density $n_\text{exc}$ within the two-chemical-potential model~\cite{Tangney2002,Murray2007,Marini2021} and yields the photoexcited phonon spectrum $\omega^1_{\mathbf{q}\nu}$, while the ground-state calculation yields $\omega^0_{\mathbf{q}\nu}$. 
Once both sets of inputs are available, the phonon-coherence calculation can be performed entirely within \textsc{EPI}q.
 
\subsection{Wannier interpolation of the electron-phonon matrix elements}
\label{sec:wannier}

Evaluating the reciprocal-space integrals of Eq.~\eqref{eq:eom-aa} on grids fine enough to converge the delta functions makes the direct \emph{ab initio} evaluation of $g^\nu_{mn}(\mathbf{k},\mathbf{q})$ prohibitively expensive, so interpolation is essential. \textsc{EPI}q exploits the localization properties of maximally localized Wannier functions (MLWFs)~\cite{Marzari2012}, obtained by minimizing the spread functional over the $\mathbf{k}$-dependent unitary matrices $U_{mn}(\mathbf{k})$ that mix the Bloch states of a composite manifold. The code interfaces directly with the output of Wannier90~\cite{Pizzi2020} and Quantum ESPRESSO~\cite{Giannozzi2009,Giannozzi2017}. Matrix elements expressed in the corresponding smooth Wannier gauge vary smoothly in reciprocal space, allowing accurate interpolation onto arbitrary points in the Brillouin zone from \emph{ab initio} input calculated only on coarse grids.

Specifically, we compute the deformation-potential matrix elements using density-functional perturbation theory (DFPT)~\cite{Baroni2001},
\begin{equation}
d^{a\alpha}_{mn}(\mathbf{k}+\mathbf{q}, \mathbf{k}) =
\left\langle u_{m\mathbf{k}+\mathbf{q}} \left|
\frac{\partial v_{\mathrm{KS}}}{\partial \mathbf{u}^{a\alpha}_{\mathbf{q}}}
\right| u_{n\mathbf{k}} \right\rangle,
\label{eq:defpot-bloch}
\end{equation}
where $v_{\mathrm{KS}}$ is the Kohn-Sham potential and $\mathbf{u}^{a}_{\mathbf{q}}$ is the Fourier component of the displacement of atom $a$ along the Cartesian direction $\alpha$. The electron-phonon coupling at arbitrary $\mathbf{k}$ and $\mathbf{q}$ can then be reconstructed using either conventional Fourier Wannier interpolation~\cite{PhysRevB.76.165108,Calandra2010,Giustino2017,Marini2024} or direct Wannier interpolation in the optimally smooth subspace~\cite{Volpato2025}.

\subsection{Workflow for the phonon-coherence dynamics}
\label{sec:workflow}
 
The code reads the Wannier electronic band structure together with the ground-state and photoexcited deformation-potential matrix elements and interpolates the electronic structure onto the fine grids entering Eq.~\eqref{eq:eom-aa}. We use the photoexcited Wannier-interpolated band structure to compute the Kohn-Sham eigenvalues $\epsilon^1_{n\mathbf{k}}$ entering the reciprocal-space integrals that determine $F_{\mathbf{q}\nu}(t)$ and $\Gamma_{\mathbf{q}\nu}(t)$ in Eq.~\eqref{eq:structure}. This choice is motivated by the observation that, within the quasi-equilibrium photoexcited picture, the dominant contributions to Eq.~\eqref{eq:eom-aa} arise from electronic transitions close to the valence and conduction quasi-Fermi levels, owing to the energy-conservation constraints imposed by the delta functions. Provided that photoexcitation does not qualitatively modify the relevant band dispersion, i.e. that its main effect is a redistribution of the occupations, the essential requirement is an accurate description of the electronic states satisfying these resonance conditions. Under these assumptions, the calculation can be restricted to a subset of Wannierized bands, provided that all bands close to the two quasi-Fermi levels are retained.

At each target $\mathbf{q}$ point, the phonon dispersions are interpolated for both configurations, yielding the ground-state and photoexcited frequencies, $\omega^0_{\mathbf q\nu}$ and $\omega^1_{\mathbf q\nu}$, for all phonon branches. The code then initializes the electronic and phononic occupations for both configurations. Let $N_v$ be the number of valence bands, and let $\mu_v$ and $\mu_c$ denote the two quasi-Fermi levels of the two-chemical-potential model at electronic temperature $T_\text{el}$. The electronic occupations are initialized as
\begin{align}
\nonumber f^0_{n\mathbf{k}} = 1, \quad f^1_{n\mathbf{k}} &= \left( e^{(\epsilon_{n\mathbf{k}}^1 - \mu_v)/k_BT_\text{el}} + 1 \right)^{-1} \quad \text{for } n \le N_v, \\
f^0_{n\mathbf{k}} = 0, \quad f^1_{n\mathbf{k}} &= \left( e^{(\epsilon_{n\mathbf{k}}^1 - \mu_c)/k_BT_\text{el}} + 1 \right)^{-1} \quad \text{otherwise},
\label{eq:init-el}
\end{align}
where the superscripts $0$ and $1$ denote the ground state and the quasi-equilibrium photoexcited state reached after intraband carrier equilibration, respectively.

The phonon occupations are initialized as thermal Bose-Einstein occupations at temperature $T_0$. Throughout the present implementation, they are assumed to remain described by a thermal distribution characterized by a time-dependent temperature $T(t)$, whose evolution is specified in the next subsection.

 \begin{figure*}[t]
  \includegraphics[width=0.99\textwidth]{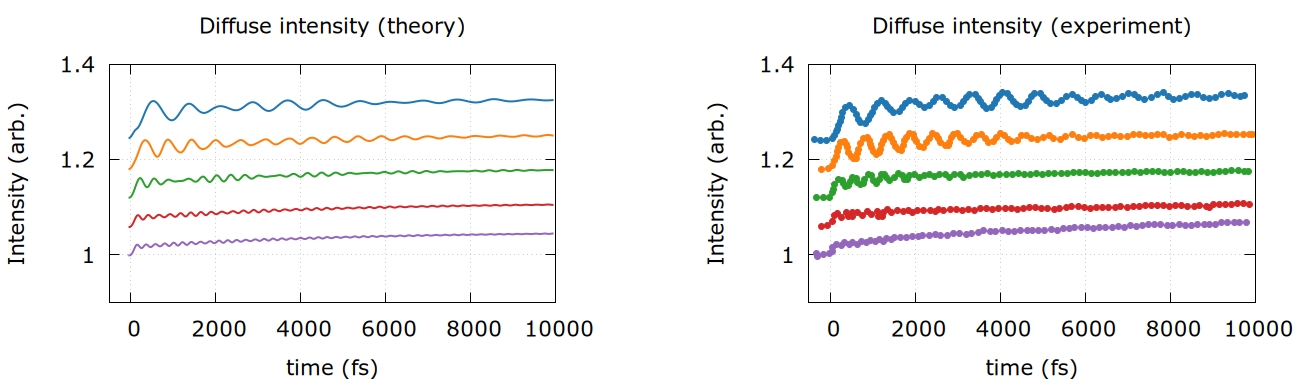}
  \caption{Calculated (left) and measured (right) time-resolved diffuse
  intensity of photoexcited germanium at the five momentum transfers of Fig.~1 in Ref.~\onlinecite{Zhu2015}. Curves are normalized to the pre-laser intensity
  and vertically offset for clarity. Calculations are performed at
  $n_\text{exc}=0.0075/\text{primitive cell}$ and $T_\text{el}=1000$~K.}
  \label{fig:traces}
\end{figure*}

\subsection{Time evolution of the occupations}
\label{sec:time-evolution}
 
Because the phonon-coherence dynamics of interest develops over timescales much longer than the photoexcitation process itself, the present implementation models the time evolution of the electronic and phononic occupations phenomenologically, leaving full many-body dynamics~\cite{Mocatti2026} to future work. Within this approach, three ingredients capture the expected photoinduced evolution: a transition of the electronic occupations from their ground-state values to their photoexcited values over a few hundred femtoseconds, consistent with observed and predicted intraband carrier-thermalization times; a smooth evolution of the phonon frequencies over a comparable timescale, associated with the onset of the two-distribution regime; and a slower evolution of the phonon occupations over picosecond timescales, reflecting both the modification of the vibrational spectrum and, to a large extent, the gradual transfer of energy to the lattice. Accordingly, at each time step, the electronic occupations and phonon frequencies are computed as
\begin{align}
\nonumber f_{n\mathbf{k}}(t) &= f^0_{n\mathbf{k}} + \left( f^1_{n\mathbf{k}} - f^0_{n\mathbf{k}} \right) \theta(t, \Delta t_\text{el}), \\
\omega_{\mathbf{q}\nu}(t) &= \omega^0_{\mathbf{q}\nu} + \left( \omega^1_{\mathbf{q}\nu} - \omega^0_{\mathbf{q}\nu} \right) \theta(t, \Delta t_\text{ph}),
\label{eq:switching}
\end{align}
where $t=0$ denotes the arrival time of the photoexcitation pulse and $\theta(t,\Delta t)$ is a smooth switching function evolving from $0$ to $1$ over the interval $\Delta t$. The specific short-time interpolation is not expected to affect the long-time dynamics significantly, provided that it ensures a smooth transition between the two reference states. In this work, we set both $\Delta t_\text{el}$ and $\Delta t_\text{ph}$ to $200$~fs as representative short-time relaxation scales. We employ the following smoothstep function
\begin{equation}
\theta(t,\Delta t)=
\begin{cases}
0, & t\leq 0,\\[3pt]
y(t/\Delta t),
& 0<t<\Delta t,\\[6pt]
1, & t\geq \Delta t,
\end{cases}
\label{eq:smoothstep}
\end{equation}
with 
\begin{equation}
    y(x) = 6 x^5
-15 x^4
+10 x^3.
\end{equation}
The logarithmic derivative $\dot{\omega}_{\mathbf{q}\nu}/\omega_{\mathbf{q}\nu}$ resulting from Eq.~\eqref{eq:switching} enters directly the squeezing drive of Eq.~\eqref{eq:eom-aa}. The phonon occupations are evolved by imposing a slow exponential heating of the lattice:
\begin{equation}
\begin{gathered}
n_{\mathbf{q}\nu}(t) =
\left( e^{\beta(t)\hbar\omega_{\mathbf{q}\nu}(t)} - 1 \right)^{-1}, \\
\beta(t) =
\frac{1}{
k_B\left[
T_0+\Delta T\left(1-e^{-t/\tau_\text{ph}}\right)
\right]},
\label{eq:lattice-heating}
\end{gathered}
\end{equation}
where $\Delta T$ is the lattice-temperature increase following photoexcitation and $\tau_\text{ph}$ is the characteristic timescale of energy transfer to the lattice. The electronic temperature is brought to the lattice temperature over the same timescale, reflecting the expected long-time thermalization of the two subsystems and ensuring that the driving term $F_{\mathbf{q}\nu}(t)$ vanishes at quasi-equilibrium~\cite{Stefanucci2024}. The asymptotic value $T_0+\Delta T$ defines the long-time lattice temperature, whose effect on the simulated dynamics is examined in Appendix~\ref{app:comparison-lattice-temperature}.
 
\subsection{Numerical aspects}
\label{sec:numerics}
 
Once $f_{n\mathbf{k}}(t)$, $\omega_{\mathbf{q}\nu}(t)$, and $n_{\mathbf{q}\nu}(t)$ have been updated, the code evaluates the time-dependent driving and damping terms of Eq.~\eqref{eq:structure} and propagates Eq.~\eqref{eq:eom-aa} in time. The numerical propagation uses the Dormand--Prince RKDP54~\cite{Dormand1980} scheme already implemented in \textsc{EPI}q for the electron-phonon Bloch equations without anomalous phonon correlators~\cite{Mocatti2026}.

Fine electronic interpolation is essential to converge the reciprocal-space integrals. The delta functions in Eq.~\eqref{eq:eom-aa} are represented by smearing functions of width $\sigma$, and all observables are computed simultaneously for a range of $\sigma$ values, allowing convergence with respect to the smearing width and the electronic $\mathbf{k}$- and phonon $\mathbf{q}$-mesh densities to be monitored within a single run.

Anharmonic and defect-induced broadening, which is not captured by the electron-phonon damping rate $\Gamma_{\mathbf{q}\nu}$, can be included through an additional phenomenological mode-resolved damping rate, whose effect is assessed in Appendix~\ref{app:comparison-anharmonic}. At each time step, the code evaluates Eq.~\eqref{eq:intensity} for a list of requested momentum transfers, using Cromer--Mann atomic form factors for X-ray probes~\cite{Cromer1968} and their Mott--Bethe transformation for electron probes. This allows both XFEL and UED observables to be obtained from the same simulation.

\section{Results}
\label{sec:results}

We apply the framework to photoexcited germanium, for which high-quality
time-resolved diffuse scattering data are available from Ref.~\onlinecite{Zhu2015}. In that experiment, a Ge(100) crystal was pumped at
\SI{800}{\nano\meter} and probed at \SI{10.2}{\kilo\electronvolt} in a grazing-incidence geometry, with the diffuse intensity monitored over a
\SI{10}{\pico\second} delay window.

Ground-state and photoexcited calculations were performed within density-functional perturbation theory, and all computational details are given in Appendix~\ref{app:computational}. The photoexcited electronic distribution was modelled using two quasi-Fermi levels at a common carrier temperature~\cite{Marini2021}, with the excited-carrier density $n_\text{exc}$ and electronic temperature $T_\text{el}$ defining the photoexcited occupations.

\begin{figure}[t]
  \centering
  \includegraphics[width=0.99\columnwidth]{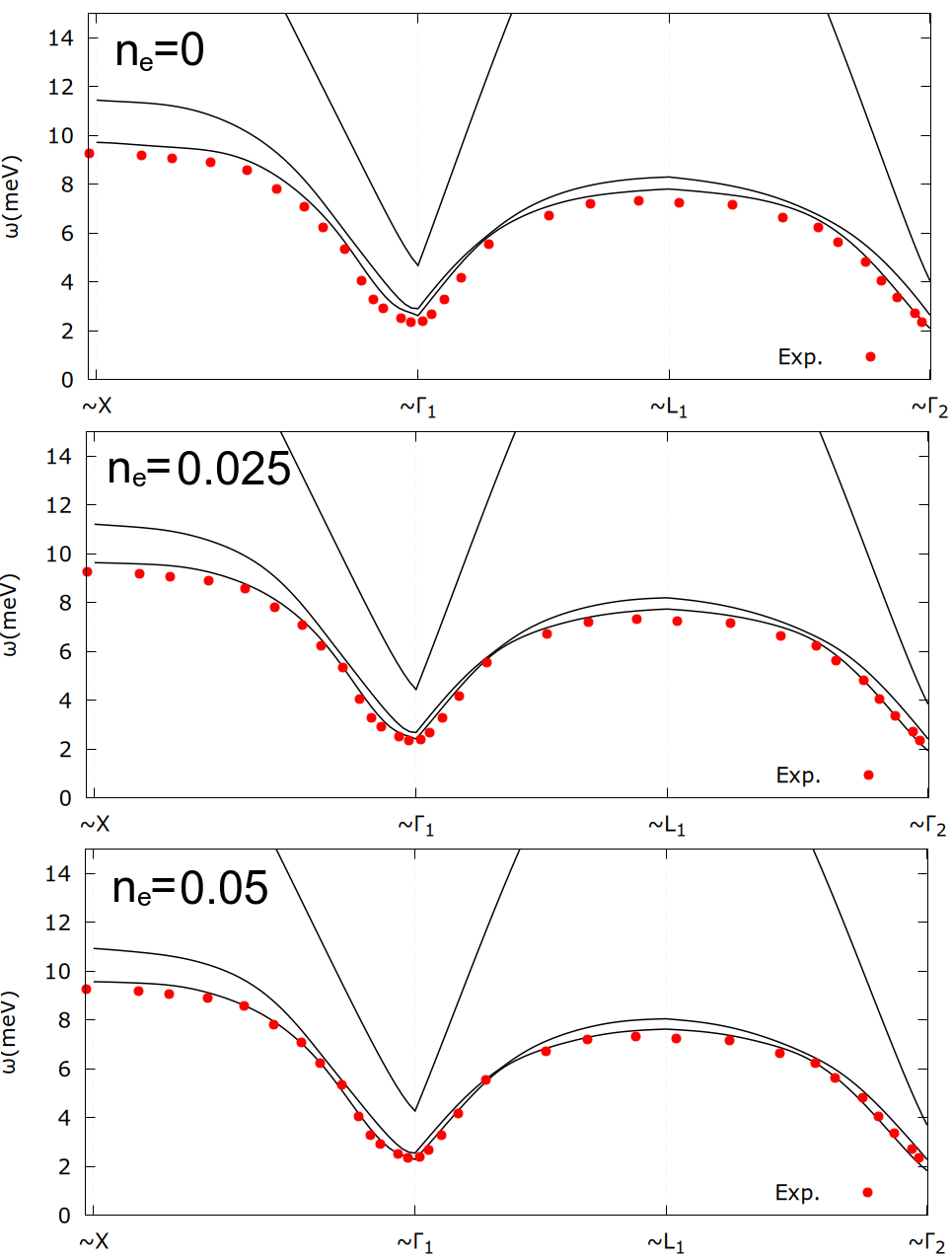}
  \caption{Calculated phonon dispersion of germanium along the reciprocal-space
  path probed in Ref.~\onlinecite{Zhu2015} (lines), compared with the
  frequencies extracted from the Fourier transform of the measured diffuse
  intensity (symbols). Curves are shown for the ground state and for excited-carrier densities of $0.025$ and $0.05$ electrons per primitive cell. High-symmetry
  labels carry a tilde because the experimental cut passes near, but not
  through, the corresponding points.}
  \label{fig:dispersion}
\end{figure}

\subsection{Time traces of the diffuse intensity}

Figure~\ref{fig:traces} compares the calculated diffuse intensity at $T_\text{el}=1000$~K and $n_\text{exc}=0.0075/\text{primitive cell}$ with the measured traces at the five momentum transfers of Ref.~\onlinecite{Zhu2015}. In addition, a phenomenological mode- and momentum-independent damping $\Gamma_\text{phen}=0.15$~meV is included to account for the experimentally observed decay of the coherent oscillations arising from defects and anharmonicity, which are not captured by the present treatment. This value is slightly larger than the calculated anharmonic phonon linewidths of Ge reported in Fig.~3 of Ref.~\onlinecite{PhysRevB.67.144304}. Its effect is further examined in Appendix~\ref{app:comparison-anharmonic}.

Overall, several key experimental features are reproduced by the theoretical approach. First, the oscillation frequencies are reproduced across the full set of momentum transfers and are set by $2\omega_{\mathbf{q}\nu}$, obtained directly from the two-chemical-potential calculation. Second, the oscillation amplitude decreases with increasing $|\mathbf{q}|$, in qualitative agreement with experiment, reflecting the combined momentum dependence of the prefactor and mode-resolved matrix elements entering Eq.~\eqref{eq:intensity}. Third, the monotonically rising incoherent background is reproduced. This component is absent from sudden-quench models, which predict a purely oscillatory signal about a constant offset; here it arises from the increase of $n_{\mathbf{q}\nu}(t)$ in Eq.~\eqref{eq:intensity} as energy is transferred from the photoexcited carriers to the lattice. A good agreement with experiment is obtained by assuming a lattice-temperature increase of $\Delta T=10$~K with a characteristic timescale $\tau_\text{ph}=5$~ps. A more detailed comparison between theory and experiment is presented in Appendix~\ref{app:comparison}, where the two sets of curves are superimposed. The reference calculation uses $T_\text{el}=1000$~K and an asymptotic lattice temperature of $310$~K.

In Fig.~\ref{fig:dispersion}, we compare the transverse-acoustic phonon frequencies extracted from experiment with those predicted by the two-chemical-potential approach at different photoexcitation densities. The transverse-acoustic branches soften monotonically with increasing $n_\text{exc}$, as expected from the weakening of the covalent bonds upon promotion of electrons from bonding to antibonding states. The softening is strongly wavevector dependent and is largest near the zone boundary. The relatively weak dependence of the acoustic branches on photoexcitation makes a precise determination of the photocarrier concentration difficult. The comparison with the experimental frequencies in Fig.~\ref{fig:dispersion} is consistent with excitation densities in the range $n_\text{exc}=0$--$0.05$ electrons per primitive cell, including the value $n_\text{exc}=0.0075$ used to reproduce the time-dependent diffuse intensity.

\section{Discussion}
\label{sec:discussion}

We now discuss the main approximations underlying the results. First, the no-mode-mixing assumption is well satisfied in Ge, where the photoexcited eigenvectors are essentially unchanged from their ground-state counterparts, but is expected to fail in systems undergoing a symmetry-lowering structural transition, for which the general Eqs.~\eqref{eq:eom-mixing-1}-\eqref{eq:eom-mixing-2} must be used. VO$_2$ represents an interesting future test case.

Second, the adiabatic separation underlying Eq.~\eqref{eq:hamiltonian} requires the carrier distribution to evolve slowly compared with the electronic dephasing, but not necessarily compared with the phonon periods. In the first few hundred femtoseconds after excitation, these assumptions are not fully justified, motivating the need for a full many-body treatment.

A third limitation of the present approach, identified empirically in Appendix~\ref{app:comparison-overlay}, is the use of a single time-dependent lattice temperature to populate all phonon modes. Indeed, the five traces of Fig.~\ref{fig:traces} suggest that the post-excitation phonon population is not thermal on the timescale of the measurement, with different $\mathbf{q}\nu$ modes exhibiting different incoherent-background growth. Reproducing this behaviour requires propagating $n_{\mathbf{q}\nu}(t)$ beyond the assumption of a common thermal distribution.

The quasi-thermal treatment of the electronic and phononic occupations shares a common origin: their real-time evolution is supplied phenomenologically rather than computed microscopically. The electron-phonon and phonon-phonon collision integrals required for this purpose have recently been implemented in \textsc{EPI}q~\cite{Mocatti2026}, and the two implementations will be integrated in future work. Propagating $f_{n\mathbf{k}}(t)$ and $n_{\mathbf{q}\nu}(t)$ self-consistently alongside $\langle \hat{a}_{\mathbf{q}\nu}\hat{a}_{-\mathbf{q}\nu}\rangle$ would replace the switching function of Eq.~\eqref{eq:switching} and the common-$T(t)$ occupations of Sec.~\ref{sec:implementation} with nonequilibrium, $\mathbf{q}\nu$-resolved distributions, while also providing a microscopic description of anharmonic decay beyond the phenomenological $\Gamma_\text{phen}$ introduced in Appendix~\ref{app:comparison-anharmonic}. We expect this to improve the description of the over- and underestimated background growth documented in Appendix~\ref{app:comparison-overlay}, and leave its implementation to future work.

\section{Conclusions}

We have formulated the time-resolved first-order diffuse scattering intensity from first principles within the density-matrix formalism, identifying the anomalous phonon correlator $\langle \hat{a}_{\mathbf{q}\nu}\hat{a}_{-\mathbf{q}\nu}\rangle$ as the source of the coherent contribution and the phonon occupations as the origin of the incoherent one. We have further shown that the anomalous correlations are driven both by the photoinduced frequency shift and by electron-phonon scattering off the nonequilibrium carrier distribution. The scheme has been implemented in \textsc{EPI}q and validated against femtosecond diffuse scattering measurements on germanium, reproducing the oscillation frequencies, their wavevector dependence, and the rising incoherent background in good agreement with experiment. The framework therefore provides a route to the quantitative interpretation of ultrafast diffuse scattering, including in systems where mode mixing or structural instabilities render simple models inadequate.

The principal remaining approximation is the phenomenological treatment of the real-time evolution of the electronic and phononic populations, which the comparison with experiment identifies as a major source of the residual discrepancies. Integrating the present scheme with the first-principles carrier and phonon dynamics implemented in \textsc{EPI}q~\cite{Mocatti2026} would remove the need for this phenomenological treatment and will be the subject of future work.

\section*{Data Availability}
The data supporting the findings of this study are available in the Zenodo repository (link to be inserted). 

\section*{Acknowledgments}
This work was funded/co-funded by the European Union (ERC, DELIGHT, 101052708). Views and opinions expressed are however those of the author(s) only and do not necessarily reflect those of the European Union or the European Research Council. Neither the European Union nor the granting authority can be held responsible for them.

\appendix
\section{Sudden-quench model and mode mixing}
\label{app:sudden}

Oscillations of the diffuse intensity away from Bragg peaks can be modelled by a sudden change of the normal-mode frequencies from their ground-state to their photoexcited values~\cite{Kiss1994,Johnson2009}. For the independent-mode Hamiltonian
\begin{equation}
\hat{H}
=
\sum_{\mathbf{q}\nu}
\hbar\omega_{\mathbf{q}\nu}(t)
\left(
\hat{a}^\dagger_{\mathbf{q}\nu}
\hat{a}_{\mathbf{q}\nu}
+\frac{1}{2}
\right),
\end{equation}
undergoing an instantaneous change from $\omega^0_{\mathbf{q}\nu}$ to $\omega^1_{\mathbf{q}\nu}$ at $t_0$, continuity of the Heisenberg-picture displacements and momenta gives the Bogoliubov relations
\begin{align}
\nonumber
\hat{a}_{\mathbf{q}\nu}(t_0^+)
&=
\mathcal{U}_{\mathbf{q}\nu}
\hat{a}_{\mathbf{q}\nu}(t_0^-)
+
\mathcal{V}_{\mathbf{q}\nu}
\hat{a}_{-\mathbf{q}\nu}^\dagger(t_0^-),\\
\mathcal{U}_{\mathbf{q}\nu}
&=
\frac{1}{2}
\frac{\omega_{\mathbf{q}\nu}^0+\omega_{\mathbf{q}\nu}^1}
{\sqrt{\omega_{\mathbf{q}\nu}^0\omega_{\mathbf{q}\nu}^1}},
\qquad
\mathcal{V}_{\mathbf{q}\nu}
=
\frac{1}{2}
\frac{\omega_{\mathbf{q}\nu}^0-\omega_{\mathbf{q}\nu}^1}
{\sqrt{\omega_{\mathbf{q}\nu}^0\omega_{\mathbf{q}\nu}^1}}.
\label{eq:app-bogoliubov}
\end{align}
For an initial thermal state and assuming $\mathbf{q}\nu$ and $-\mathbf{q}\nu$ to be equivalent by symmetry, the displacement correlator becomes
\begin{widetext}
\begin{align}
\langle u_{\mathbf{q}\nu}(t)u_{-\mathbf{q}\nu}(t)\rangle
=
\frac{\hbar}{4\omega_{\mathbf{q}\nu}^0}
\left(2\langle n^0_{\mathbf{q}\nu}\rangle+1\right)
\left[
1+
\left(
\frac{\omega_{\mathbf{q}\nu}^0}
{\omega_{\mathbf{q}\nu}^1}
\right)^2
\right] +
\frac{\hbar}{4\omega_{\mathbf{q}\nu}^0}
\left(2\langle n^0_{\mathbf{q}\nu}\rangle+1\right)
\left[
1-
\left(
\frac{\omega_{\mathbf{q}\nu}^0}
{\omega_{\mathbf{q}\nu}^1}
\right)^2
\right]
\cos\left[
2\omega^1_{\mathbf{q}\nu}(t-t_0)
\right],
\label{eq:app-sudden}
\end{align}
\end{widetext}
where
\begin{equation}
\langle n^0_{\mathbf{q}\nu}\rangle
=
\left[
e^{\hbar\omega^0_{\mathbf{q}\nu}/k_BT}-1
\right]^{-1}.
\end{equation}
Equation~\eqref{eq:app-sudden} is the standard sudden-quench result recovered from Eq.~\eqref{eq:eom-aa} in the limit of an instantaneous frequency change with the electron-phonon terms switched off. A phenomenological lattice heating can additionally be included by allowing the phonon occupation entering the incoherent contribution to evolve in time.

The derivation above imposes continuity on the mode-resolved displacement and momentum coordinates across the quench and is therefore valid only when the phonon eigenvectors remain unchanged. More generally, continuity must be imposed on the atom-resolved displacements and momenta before projecting onto the photoexcited eigenvectors. This gives~\cite{Johnson2009}
\begin{equation}
\hat{a}_{\mathbf{q}\nu}(t_0^+)
=
\sum_{\nu'}
E_{\mathbf{q}\nu\nu'}
\left[
U_{\mathbf{q}\nu\nu'}
\hat{a}_{\mathbf{q}\nu'}(t_0^-)
+
V_{\mathbf{q}\nu\nu'}
\hat{a}^\dagger_{-\mathbf{q}\nu'}(t_0^-)
\right],
\end{equation}
together with the analogous relations for the remaining operators, where
\begin{align}
\nonumber
E_{\mathbf{q}\nu\nu'}
&=
\sum_{a\alpha}
\left(
e_{\mathbf{q}\nu}^{1a\alpha}
\right)^*
e_{\mathbf{q}\nu'}^{0a\alpha},
\\
\nonumber
U_{\mathbf{q}\nu\nu'}
&=
\frac{1}{2}
\frac{
\omega_{\mathbf{q}\nu'}^0+\omega_{\mathbf{q}\nu}^1
}{
\sqrt{
\omega_{\mathbf{q}\nu'}^0
\omega_{\mathbf{q}\nu}^1
}
},
\\
V_{\mathbf{q}\nu\nu'}
&=
\frac{1}{2}
\frac{
\omega_{\mathbf{q}\nu'}^0-\omega_{\mathbf{q}\nu}^1
}{
\sqrt{
\omega_{\mathbf{q}\nu'}^0
\omega_{\mathbf{q}\nu}^1
}
}.
\end{align}
Mode mixing enters through the overlap matrix $E_{\mathbf{q}\nu\nu'}$. Assuming time-reversal symmetry, the corresponding diagonal displacement correlator becomes
\begin{widetext}
\begin{align}
\nonumber
&\langle
u_{\mathbf{q}\nu}(t)
u_{-\mathbf{q}\nu}(t)
\rangle
=
\frac{\hbar}
{2\sqrt{
\omega_{\mathbf{q}\nu}^1
\omega_{-\mathbf{q}\nu}^1
}}
\Bigg[
\sum_{\nu'}
|E_{\mathbf{q}\nu\nu'}|^2
\frac{1}{2}
\frac{
\omega_{\mathbf{q}\nu}^{1\,2}
-
\omega_{\mathbf{q}\nu'}^{0\,2}
}{
\omega_{\mathbf{q}\nu'}^0
\omega_{\mathbf{q}\nu}^1
}
\left(
2\langle n^0_{\mathbf{q}\nu'}\rangle+1
\right)
\\
&\hspace{65pt}\times
\cos\left[
\left(
\omega^1_{\mathbf{q}\nu}
+
\omega^1_{-\mathbf{q}\nu}
\right)
(t-t_0)
\right]
+
\sum_{\nu'}
|E_{\mathbf{q}\nu\nu'}|^2
\frac{1}{2}
\frac{
\omega_{\mathbf{q}\nu'}^{0\,2}
+
\omega_{\mathbf{q}\nu}^{1\,2}
}{
\omega_{\mathbf{q}\nu'}^0
\omega_{\mathbf{q}\nu}^1
}
\left(
2\langle n^0_{\mathbf{q}\nu'}\rangle+1
\right)
\Bigg].
\end{align}
\end{widetext}
Because the photoexcited normal modes are linear combinations of the ground-state ones, expectation values over the initial thermal state are generally not diagonal in the final-mode basis. The fully general correlator
$\langle u_{\mathbf{q}\nu}u_{-\mathbf{q}\bar{\nu}}\rangle$
therefore contains cross terms oscillating at both
$\omega^1_{\mathbf{q}\nu}+\omega^1_{\mathbf{q}\bar{\nu}}$
and
$\omega^1_{\mathbf{q}\nu}-\omega^1_{\mathbf{q}\bar{\nu}}$.

\section{Closure of the hierarchy}
\label{app:closure}

Applying the Heisenberg equation to the correlators entering Eq.~\eqref{eq:target} generates two relevant three-point correlators,
\[
S_{mn}^\nu(\mathbf{k},\mathbf{q})
=
\left\langle
\hat{c}^\dagger_{m\mathbf{k}+\mathbf{q}}
\hat{c}_{n\mathbf{k}}
\hat{a}_{\mathbf{q}\nu}
\right\rangle
\]
and
\[
R_{mn}^\nu(\mathbf{k},\mathbf{q})
=
\left\langle
\hat{c}^\dagger_{m\mathbf{k}-\mathbf{q}}
\hat{c}_{n\mathbf{k}}
\hat{a}^\dagger_{\mathbf{q}\nu}
\right\rangle,
\]
while the remaining terms follow by relabelling
$\nu\leftrightarrow\nu'$ and
$\mathbf{q}\leftrightarrow-\mathbf{q}$.
For $\mathbf{q}\neq\Gamma$, their equations of motion are

\begin{widetext}
\begin{align}
\nonumber
\dot{S}_{mn}^\nu(t,\mathbf{k},\mathbf{q})
={}&
\frac{i}{\hbar}
\left(
\epsilon_{m\mathbf{k}+\mathbf{q}}
-
\epsilon_{n\mathbf{k}}
-
\hbar\omega_{\mathbf{q}\nu}(t)
\right)
S_{mn}^\nu(t,\mathbf{k},\mathbf{q})
+
\mathcal{O}\!\left(\{\dot{e}^{a\alpha}_{\mathbf{q}\nu}\}\right)
\\
\nonumber
&-
\frac{i}{\hbar}
\sum_{\bar n\bar{\mathbf q}\bar\nu}
g_{n\bar n}^{\bar\nu}
(t,\mathbf{k}-\bar{\mathbf q},\bar{\mathbf q})
\Big(
\langle
\hat c^\dagger_{m\mathbf{k}+\mathbf q}
\hat c_{\bar n\mathbf{k}-\bar{\mathbf q}}
\hat a_{\mathbf q\nu}
\hat a_{\bar{\mathbf q}\bar\nu}
\rangle
+
\langle
\hat c^\dagger_{m\mathbf{k}+\mathbf q}
\hat c_{\bar n\mathbf{k}-\bar{\mathbf q}}
\hat a_{\mathbf q\nu}
\hat a^\dagger_{-\bar{\mathbf q}\bar\nu}
\rangle
\Big)
\\
\nonumber
&+
\frac{i}{\hbar}
\sum_{\bar m\bar{\mathbf q}\bar\nu}
g_{\bar m n}^{\bar\nu}
(t,\mathbf{k}+\mathbf q,\bar{\mathbf q})
\Big(
\langle
\hat c^\dagger_{\bar m\mathbf{k}+\mathbf q+\bar{\mathbf q}}
\hat c_{n\mathbf{k}}
\hat a_{\mathbf q\nu}
\hat a_{\bar{\mathbf q}\bar\nu}
\rangle
+
\langle
\hat c^\dagger_{\bar m\mathbf{k}+\mathbf q+\bar{\mathbf q}}
\hat c_{n\mathbf{k}}
\hat a_{\mathbf q\nu}
\hat a^\dagger_{-\bar{\mathbf q}\bar\nu}
\rangle
\Big)
\\
&-
\frac{i}{\hbar}
\sum_{\bar n\bar m\bar{\mathbf k}}
g^\nu_{\bar m\bar n}
(t,\bar{\mathbf k},-\mathbf q)
\left\langle
\hat c^\dagger_{\bar m\bar{\mathbf k}-\mathbf q}
\hat c_{\bar n\bar{\mathbf k}}
\hat c^\dagger_{m\mathbf{k}+\mathbf q}
\hat c_{n\mathbf{k}}
\right\rangle .
\label{eq:app-Sdot}
\end{align}

\begin{align}
\nonumber
\dot{R}_{mn}^\nu(t,\mathbf{k},\mathbf{q})
={}&
\frac{i}{\hbar}
\left(
\epsilon_{m\mathbf{k}-\mathbf q}
-
\epsilon_{n\mathbf{k}}
+
\hbar\omega_{\mathbf q\nu}(t)
\right)
R_{mn}^\nu(t,\mathbf{k},\mathbf{q})
+
\mathcal{O}\!\left(\{\dot e^{a\alpha}_{\mathbf q\nu}\}\right)
\\
\nonumber
&-
\frac{i}{\hbar}
\sum_{\bar n\bar{\mathbf q}\bar\nu}
g_{n\bar n}^{\bar\nu}
(t,\mathbf{k}-\bar{\mathbf q},\bar{\mathbf q})
\Big(
\langle
\hat c^\dagger_{m\mathbf{k}-\mathbf q}
\hat c_{\bar n\mathbf{k}-\bar{\mathbf q}}
\hat a^\dagger_{\mathbf q\nu}
\hat a_{\bar{\mathbf q}\bar\nu}
\rangle
+
\langle
\hat c^\dagger_{m\mathbf{k}-\mathbf q}
\hat c_{\bar n\mathbf{k}-\bar{\mathbf q}}
\hat a^\dagger_{\mathbf q\nu}
\hat a^\dagger_{-\bar{\mathbf q}\bar\nu}
\rangle
\Big)
\\
\nonumber
&+
\frac{i}{\hbar}
\sum_{\bar m\bar{\mathbf q}\bar\nu}
g_{\bar m n}^{\bar\nu}
(t,\mathbf{k}-\mathbf q,\bar{\mathbf q})
\Big(
\langle
\hat c^\dagger_{\bar m\mathbf{k}-\mathbf q+\bar{\mathbf q}}
\hat c_{n\mathbf{k}}
\hat a^\dagger_{\mathbf q\nu}
\hat a_{\bar{\mathbf q}\bar\nu}
\rangle
+
\langle
\hat c^\dagger_{\bar m\mathbf{k}-\mathbf q+\bar{\mathbf q}}
\hat c_{n\mathbf{k}}
\hat a^\dagger_{\mathbf q\nu}
\hat a^\dagger_{-\bar{\mathbf q}\bar\nu}
\rangle
\Big)
\\
&+
\frac{i}{\hbar}
\sum_{\bar n\bar m\bar{\mathbf k}}
g^\nu_{\bar m\bar n}
(t,\bar{\mathbf k},\mathbf q)
\left\langle
\hat c^\dagger_{\bar m\bar{\mathbf k}+\mathbf q}
\hat c_{\bar n\bar{\mathbf k}}
\hat c^\dagger_{m\mathbf{k}-\mathbf q}
\hat c_{n\mathbf{k}}
\right\rangle .
\label{eq:app-Rdot}
\end{align}
\end{widetext}
Here, coherent-phonon renormalization of the electronic energies and terms proportional to the time derivative of the phonon frequencies have been neglected, while contributions arising from the time dependence of the phonon eigenvectors are collected in
$\mathcal{O}(\{\dot e^{a\alpha}_{\mathbf q\nu}\})$.
These terms are neglected in the subsequent Markov approximation.

The four-point correlators are factorized using a cluster decomposition~\cite{Rossi2002,Furci2026}, neglecting electronic coherences over the timescale of interest while retaining anomalous phonon correlations. For example,
\begin{align}
\nonumber
&
\left\langle
\hat c^\dagger_{m\mathbf{k}+\mathbf q}
\hat c_{\bar n\mathbf{k}-\bar{\mathbf q}}
\hat a_{\mathbf q\nu}
\hat a_{\bar{\mathbf q}\bar\nu}
\right\rangle
\\
\nonumber
&\qquad\simeq
\left\langle
\hat c^\dagger_{m\mathbf{k}+\mathbf q}
\hat c_{\bar n\mathbf{k}-\bar{\mathbf q}}
\right\rangle
\left\langle
\hat a_{\mathbf q\nu}
\hat a_{\bar{\mathbf q}\bar\nu}
\right\rangle
\\
&\qquad\simeq
f_{m\mathbf{k}+\mathbf q}(t)
\left\langle
\hat a_{\mathbf q\nu}
\hat a_{\bar{\mathbf q}\bar\nu}
\right\rangle
\delta_{m\bar n}
\delta_{\mathbf q,-\bar{\mathbf q}} .
\label{eq-ap3}
\end{align}
For the four-fermion contribution, Wick factorization gives
\begin{align}
\nonumber
&
\left\langle
\hat c^\dagger_{\bar m\bar{\mathbf k}-\mathbf q}
\hat c_{\bar n\bar{\mathbf k}}
\hat c^\dagger_{m\mathbf{k}+\mathbf q}
\hat c_{n\mathbf{k}}
\right\rangle
\\
\nonumber
&\qquad\simeq
\left\langle
\hat c^\dagger_{\bar m\bar{\mathbf k}-\mathbf q}
\hat c_{\bar n\bar{\mathbf k}}
\right\rangle
\left\langle
\hat c^\dagger_{m\mathbf{k}+\mathbf q}
\hat c_{n\mathbf{k}}
\right\rangle
\\
\nonumber
&\qquad\quad+
\left\langle
\hat c^\dagger_{\bar m\bar{\mathbf k}-\mathbf q}
\hat c_{n\mathbf{k}}
\right\rangle
\left\langle
\hat c_{\bar n\bar{\mathbf k}}
\hat c^\dagger_{m\mathbf{k}+\mathbf q}
\right\rangle
\\
&\qquad\simeq
f_{n\mathbf{k}}(t)
\left[
1-f_{m\mathbf{k}+\mathbf q}(t)
\right]
\delta_{\bar m n}
\delta_{\bar n m}
\delta_{\bar{\mathbf k},\mathbf{k}+\mathbf q},
\qquad
(\mathbf q\neq\Gamma).
\label{eq-ap4}
\end{align}
The first disconnected contraction in the second line vanishes for
$\mathbf q\neq\Gamma$.

The equations of motion for the three-point correlators then become
\begin{widetext}
\begin{align}
\nonumber
\dot{S}_{mn}^\nu(t,\mathbf{k},\mathbf q)
={}&
\frac{i}{\hbar}
\left(
\epsilon_{m\mathbf{k}+\mathbf q}
-
\epsilon_{n\mathbf{k}}
-
\hbar\omega_{\mathbf q\nu}(t)
\right)
S_{mn}^\nu(t,\mathbf{k},\mathbf q)
+
\mathcal{O}\!\left(\{\dot e^{a\alpha}_{\mathbf q\nu}\}\right)
\\
\nonumber
&-
\frac{i}{\hbar}
\sum_{\bar\nu}
g_{nm}^{\bar\nu}
(t,\mathbf{k}+\mathbf q,-\mathbf q)
\left[
f_{m\mathbf{k}+\mathbf q}(t)
-
f_{n\mathbf{k}}(t)
\right]
\left(
\langle
\hat a_{\mathbf q\nu}
\hat a_{-\mathbf q\bar\nu}
\rangle
+
\langle
\hat a_{\mathbf q\nu}
\hat a^\dagger_{\mathbf q\bar\nu}
\rangle
\right)
\\
&-
\frac{i}{\hbar}
g_{nm}^{\nu}
(t,\mathbf{k}+\mathbf q,-\mathbf q)
f_{n\mathbf{k}}(t)
\left[
1-f_{m\mathbf{k}+\mathbf q}(t)
\right].
\label{eq-ap5}
\end{align}

\begin{align}
\nonumber
\dot{R}_{mn}^\nu(t,\mathbf{k},\mathbf q)
={}&
\frac{i}{\hbar}
\left(
\epsilon_{m\mathbf{k}-\mathbf q}
-
\epsilon_{n\mathbf{k}}
+
\hbar\omega_{\mathbf q\nu}(t)
\right)
R_{mn}^\nu(t,\mathbf{k},\mathbf q)
+
\mathcal{O}\!\left(\{\dot e^{a\alpha}_{\mathbf q\nu}\}\right)
\\
\nonumber
&-
\frac{i}{\hbar}
\sum_{\bar\nu}
g_{nm}^{\bar\nu}
(t,\mathbf{k}-\mathbf q,\mathbf q)
\left[
f_{m\mathbf{k}-\mathbf q}(t)
-
f_{n\mathbf{k}}(t)
\right]
\left(
\langle
\hat a^\dagger_{\mathbf q\nu}
\hat a_{\mathbf q\bar\nu}
\rangle
+
\langle
\hat a^\dagger_{\mathbf q\nu}
\hat a^\dagger_{-\mathbf q\bar\nu}
\rangle
\right)
\\
&+
\frac{i}{\hbar}
g_{nm}^{\nu}
(t,\mathbf{k}-\mathbf q,\mathbf q)
f_{n\mathbf{k}}(t)
\left[
1-f_{m\mathbf{k}-\mathbf q}(t)
\right].
\label{eq-ap6}
\end{align}

Renormalization terms arising from a coherent phonon amplitude at $\Gamma$ have been discarded~\cite{Rossi2002}.

\section{Equations of motion for phonon correlators}
\label{app:mixing-eqs}

Without imposing the no-mode-mixing approximation, the equations of motion for the two independent correlators read
\begin{align}
\nonumber    &\frac{d}{dt} \langle \hat{a}_{\mathbf{q}\nu}(t)\hat{a}_{-\mathbf{q}\nu'}(t) \rangle = -i \left(\omega_{\mathbf{q}\nu}(t) + \omega_{-\mathbf{q}\nu'}(t)\right) \langle \hat{a}_{\mathbf{q}\nu}(t)\hat{a}_{-\mathbf{q}\nu'}(t) \rangle \\
\nonumber &+ \frac{1}{2} \frac{\dot{\omega}_{\mathbf{q}\nu}(t)}{\omega_{\mathbf{q}\nu}(t)}  \langle \hat{a}^\dagger_{-\mathbf{q}\nu}(t)\hat{a}_{-\mathbf{q}\nu'}(t)\rangle + \frac{1}{2} \frac{\dot{\omega}_{-\mathbf{q}\nu'}(t)}{\omega_{-\mathbf{q}\nu'}(t)}  \langle \hat{a}_{\mathbf{q}\nu}(t)\hat{a}^\dagger_{\mathbf{q}\nu'}(t)\rangle \\
\nonumber& +\frac{\pi}{\hbar} \sum_{mn\mathbf{k}\bar{\nu}} \left( g^\nu_{mn}(t, \mathbf{k}, \mathbf{q})\right)^*g^{\bar{\nu}}_{mn}(t, \mathbf{k}, \mathbf{q}) \Big[ \left(f_{m\mathbf{k}+\mathbf{q}}(t) - f_{n\mathbf{k}}(t) \right) \langle \hat{a}_{-\mathbf{q}\nu'} \hat{a}_{\mathbf{q}\bar{\nu}} \rangle \delta(\epsilon_{m\mathbf{k}+\mathbf{q}} - \epsilon_{n\mathbf{k}} - \hbar\omega_{\mathbf{q}\bar{\nu}}(t)) \\
\nonumber&\hspace{60pt} + \left(\left(f_{m\mathbf{k}+\mathbf{q}}(t) - f_{n\mathbf{k}}(t) \right) \langle\hat{a}_{-\mathbf{q}\nu'} \hat{a}^\dagger_{-\mathbf{q}\bar{\nu}} \rangle - f_{m \mathbf{k}+\mathbf{q}}(t) \left(1- f_{n\mathbf{k}}(t)\right) \delta_{\nu'\bar{\nu}}\right) \delta(\epsilon_{m\mathbf{k}+\mathbf{q}} - \epsilon_{n\mathbf{k}} + \hbar\omega_{-\mathbf{q}\bar{\nu}}(t))\Big] \\
\nonumber& -\frac{\pi}{\hbar} \sum_{mn\mathbf{k}\bar{\nu}}  g^{\nu'}_{mn}(t, \mathbf{k}, \mathbf{q}) \left(g^{\bar{\nu}}_{mn}(t, \mathbf{k}, \mathbf{q})\right)^* \Big[ \left(f_{m\mathbf{k}+\mathbf{q}}(t) - f_{n\mathbf{k}}(t) \right) \langle \hat{a}_{\mathbf{q}\nu} \hat{a}_{-\mathbf{q}\bar{\nu}} \rangle \delta(\epsilon_{m\mathbf{k}+\mathbf{q}} - \epsilon_{n\mathbf{k}} + \hbar\omega_{-\mathbf{q}\bar{\nu}}(t)) \\
&\hspace{60pt} + \left(\left(f_{m\mathbf{k}+\mathbf{q}}(t) - f_{n\mathbf{k}}(t) \right) \langle\hat{a}_{\mathbf{q}\nu} \hat{a}^\dagger_{\mathbf{q}\bar{\nu}} \rangle + f_{n \mathbf{k}}(t) \left(1- f_{m\mathbf{k}+\mathbf{q}}(t)\right) \delta_{\nu\bar{\nu}}\right) \delta(\epsilon_{m\mathbf{k}+\mathbf{q}} - \epsilon_{n\mathbf{k}} - \hbar\omega_{\mathbf{q}\bar{\nu}}(t))\Big],
\label{eq:eom-mixing-1}
\end{align}

\begin{align}
\nonumber  & \frac{d}{dt}\langle\hat{a}_{\mathbf{q}\nu}(t)\hat{a}^\dagger_{\mathbf{q}\nu'}(t) \rangle = -i \left(\omega_{\mathbf{q}\nu}(t) - \omega_{\mathbf{q}\nu'}(t)\right) \langle \hat{a}_{\mathbf{q}\nu}(t)\hat{a}^\dagger_{\mathbf{q}\nu'}(t) \rangle \\
\nonumber& + \frac{1}{2} \frac{\dot{\omega}_{\mathbf{q}\nu}(t)}{\omega_{\mathbf{q}\nu}(t)} \langle \hat{a}^\dagger_{-\mathbf{q}\nu}(t)\hat{a}^\dagger_{\mathbf{q}\nu'}(t) \rangle + \frac{1}{2}\frac{\dot{\omega}_{\mathbf{q}\nu'}(t)}{\omega_{\mathbf{q}\nu'}(t)} \langle \hat{a}_{\mathbf{q}\nu}(t)\hat{a}_{-\mathbf{q}\nu'}(t) \rangle \\
\nonumber& +\frac{\pi}{\hbar} \sum_{mn\mathbf{k}\bar{\nu}} \left( g^\nu_{mn}(t, \mathbf{k}, \mathbf{q})\right)^*g^{\bar{\nu}}_{mn}(t, \mathbf{k}, \mathbf{q}) \Big[ \left(f_{m\mathbf{k}+\mathbf{q}}(t) - f_{n\mathbf{k}}(t) \right) \langle \hat{a}^\dagger_{\mathbf{q}\nu'} \hat{a}^\dagger_{-\mathbf{q}\bar{\nu}} \rangle \delta(\epsilon_{m\mathbf{k}+\mathbf{q}} - \epsilon_{n\mathbf{k}} - \hbar\omega_{\mathbf{q}\bar{\nu}}(t)) \\
\nonumber&\hspace{60pt} + \left(\left(f_{m\mathbf{k}+\mathbf{q}}(t) - f_{n\mathbf{k}}(t) \right) \langle\hat{a}^\dagger_{\mathbf{q}\nu'} \hat{a}_{\mathbf{q}\bar{\nu}} \rangle + f_{m \mathbf{k}+\mathbf{q}}(t) \left(1- f_{n\mathbf{k}}(t)\right) \delta_{\nu'\bar{\nu}}\right) \delta(\epsilon_{m\mathbf{k}+\mathbf{q}} - \epsilon_{n\mathbf{k}} + \hbar\omega_{-\mathbf{q}\bar{\nu}}(t))\Big] \\
\nonumber& +\frac{\pi}{\hbar} \sum_{mn\mathbf{k}\bar{\nu}}  g^{\nu'}_{mn}(t, \mathbf{k}, \mathbf{q}) \left(g^{\bar{\nu}}_{mn}(t, \mathbf{k}, \mathbf{q})\right)^* \Big[ \left(f_{m\mathbf{k}+\mathbf{q}}(t) - f_{n\mathbf{k}}(t) \right) \langle \hat{a}_{\mathbf{q}\nu} \hat{a}_{-\mathbf{q}\bar{\nu}} \rangle \delta(\epsilon_{m\mathbf{k}+\mathbf{q}} - \epsilon_{n\mathbf{k}} + \hbar\omega_{-\mathbf{q}\bar{\nu}}(t)) \\
&\hspace{60pt} + \left(\left(f_{m\mathbf{k}+\mathbf{q}}(t) - f_{n\mathbf{k}}(t) \right) \langle\hat{a}_{\mathbf{q}\nu} \hat{a}^\dagger_{\mathbf{q}\bar{\nu}} \rangle + f_{n \mathbf{k}}(t) \left(1- f_{m\mathbf{k}+\mathbf{q}}(t)\right) \delta_{\nu\bar{\nu}}\right) \delta(\epsilon_{m\mathbf{k}+\mathbf{q}} - \epsilon_{n\mathbf{k}} - \hbar\omega_{\mathbf{q}\bar{\nu}}(t))\Big].
\label{eq:eom-mixing-2}
\end{align}
\end{widetext}
In the no-mode-mixing limit, Eq.~\eqref{eq:eom-mixing-1} reduces to Eq.~\eqref{eq:eom-aa}, while Eq.~\eqref{eq:eom-mixing-2} reduces to Eq.~\eqref{eq:eom-adag}.%


\section{Effect of coherent phonon at $\Gamma$}
\label{app:bragg}

The second term of Eq.~\eqref{eq:expansion} involves the expectation value of the ionic displacement at $\Gamma$, resulting in coherent Bragg-peak oscillations. This expectation value also appears in the equations of motion for coherences and occupations through the cluster decomposition of four-point correlators. We neglect it here, but it should be retained in the presence of a coherent phonon. At equilibrium its contribution vanishes, but this is not necessarily the case following optical pumping. We discuss this case within a simple model, starting from the single-mode electron-phonon Hamiltonian
\begin{equation}
\begin{gathered}
    \hat{H}_{\mathbf{q}\nu} = \hbar\omega_{\mathbf{q}\nu} \hat{a}^\dagger_{\mathbf{q}\nu} \hat{a}_{\mathbf{q}\nu} + \\
    \frac{1}{\sqrt{N}} \sum_{mn\mathbf{k}} g_{mn\nu}(\mathbf{k}, \mathbf{q}) \hat{c}^\dagger_{m\mathbf{k}+\mathbf{q}} \hat{c}_{n\mathbf{k}} (\hat{a}_{\mathbf{q}\nu} + \hat{a}^\dagger_{-\mathbf{q}\nu}),
    \label{eq:app-H}
\end{gathered}
\end{equation}
where the free electronic Hamiltonian has been omitted since it does not contribute directly to the equation of motion for the phonon operator. The Heisenberg equation of motion for $\hat{a}_{\mathbf{q}\nu}$ gives
\begin{equation}
\begin{gathered}
    i\hbar \frac{d}{dt}\langle \hat{a}_{\mathbf{q}\nu}\rangle_{\hat{\rho}_e(t)}
    = \hbar\omega_{\mathbf{q}\nu}\langle \hat{a}_{\mathbf{q}\nu}\rangle_{\hat{\rho}_e(t)} + \\
    \frac{1}{\sqrt{N}} \sum_{mn\mathbf{k}} g_{mn\nu}(\mathbf{k}, \mathbf{q}) \langle\hat{c}^\dagger_{m\mathbf{k}+\mathbf{q}} \hat{c}_{n\mathbf{k}} \rangle_{\hat{\rho}_e(t)},
\label{eq:app-eom}
\end{gathered}
\end{equation}
provided that the full state factorizes as $\hat{\rho} =
\hat{\rho}_\text{ions}(t) \otimes \hat{\rho}_e(t)$, as expected when the
electronic and ionic timescales can be separated. Formal integration gives
\begin{equation}
\begin{gathered}
    \langle\hat{a}_{\mathbf{q}\nu}(t)\rangle = \langle\hat{a}_{\mathbf{q}\nu}(0)\rangle e^{-i\omega_{\mathbf{q}\nu}t} \\
    - \frac{i}{\hbar} \int_0^t \frac{1}{\sqrt{N}} \sum_{mn\mathbf{k}} g_{mn\nu}(\mathbf{k}, \mathbf{q}) \langle\hat{c}^\dagger_{m\mathbf{k}+\mathbf{q}} \hat{c}_{n\mathbf{k}} \rangle_{\hat{\rho}_e(t')} e^{-i\omega_{\mathbf{q}\nu}(t-t')} \, dt',
    \label{eq:app-integrated}
\end{gathered}
\end{equation}
showing that the time-dependent electronic distribution provides a driving term for the coherent phonon.

Assuming spatially uniform optical pumping, crystal momentum is conserved. Neglecting electronic coherences after the initial dephasing, one has
$\langle\hat{c}^\dagger_{m\mathbf{k}+\mathbf{q}} \hat{c}_{n\mathbf{k}}
\rangle \propto \delta_{\mathbf{q}0}\delta_{mn}$, so that $\mathbf{q}$ must be $\Gamma$. Parametrizing the population relaxation as
$\langle\hat{c}^\dagger_{n\mathbf{k}}
\hat{c}_{n\mathbf{k}} \rangle_t = (f_{n\mathbf{k}}^\text{photo} -
f_{n\mathbf{k}}^0)e^{-t/\tau_e} + f_{n\mathbf{k}}^0$, the equilibrium contribution is already balanced at the ground-state equilibrium geometry, and only the photoinduced change drives the coherent displacement. Denoting this contribution by $\beta_{\Gamma\nu}(t)$ gives
\begin{widetext}
\begin{equation}
    \beta_{\Gamma\nu}(t) =
    \frac{i}{\hbar\sqrt{N}} \sum_{n\mathbf{k}} g_{nn\nu}(\mathbf{k}, \mathbf{0})
    (f_{n\mathbf{k}}^\text{photo} - f_{n\mathbf{k}}^0)
    \frac{\tau_e}{1-i\omega_{\Gamma\nu}\tau_e}
    \left(e^{(i\omega_{\Gamma\nu}-1/\tau_e)t} - 1\right)
    e^{-i\omega_{\Gamma\nu}t}.
\label{eq:app-beta}
\end{equation}
\end{widetext}
For a mode initially in a thermal state, the Glauber--Sudarshan $P$
representation~\cite{GlauberPhysRev.131.2766,SudarshanPhysRevLett.10.277},
\begin{equation}
    \hat{\rho}_\text{th} = \int \frac{1}{\pi \langle n_{\mathbf{q}\nu}\rangle} e^{-|\alpha|^2 / \langle n_{\mathbf{q}\nu} \rangle} \ket{\alpha}\bra{\alpha} \, d^2\alpha,
\label{eq:app-P}
\end{equation}
allows the driven state to be written as
$\hat{\rho}_\text{th}(t) = \int \pi^{-1}\langle n\rangle^{-1} e^{-|\alpha|^2 /
\langle n \rangle} \ket{\alpha + \beta(t)}\bra{\alpha + \beta(t)} \,
d^2\alpha$, from which
\begin{align}
\nonumber \langle \hat{a}_{\mathbf{q}\nu}^\dagger \hat{a}_{\mathbf{q}\nu}\rangle &= \langle n_{\mathbf{q}\nu} \rangle + \left| \beta_{\mathbf{q}\nu}(t)\right|^2, \\
\nonumber \langle  \hat{a}_{\mathbf{q}\nu}\rangle &= \beta_{\mathbf{q}\nu}(t), \\
\langle  \hat{a}_{\mathbf{q}\nu} \hat{a}_{\mathbf{q}\nu}\rangle &= \beta_{\mathbf{q}\nu}^2(t).
\label{eq:app-moments}
\end{align}
The driven thermal state therefore acquires a coherent displacement, together with the associated contribution $|\beta_{\mathbf{q}\nu}(t)|^2$ to the phonon population and a nonzero anomalous correlator $\beta_{\mathbf{q}\nu}^2(t)$. Thus, in systems where coherent phonons at $\Gamma$ are activated, these contributions must be retained explicitly.

\begin{figure}[t]
  \centering
  \includegraphics[width=\columnwidth]{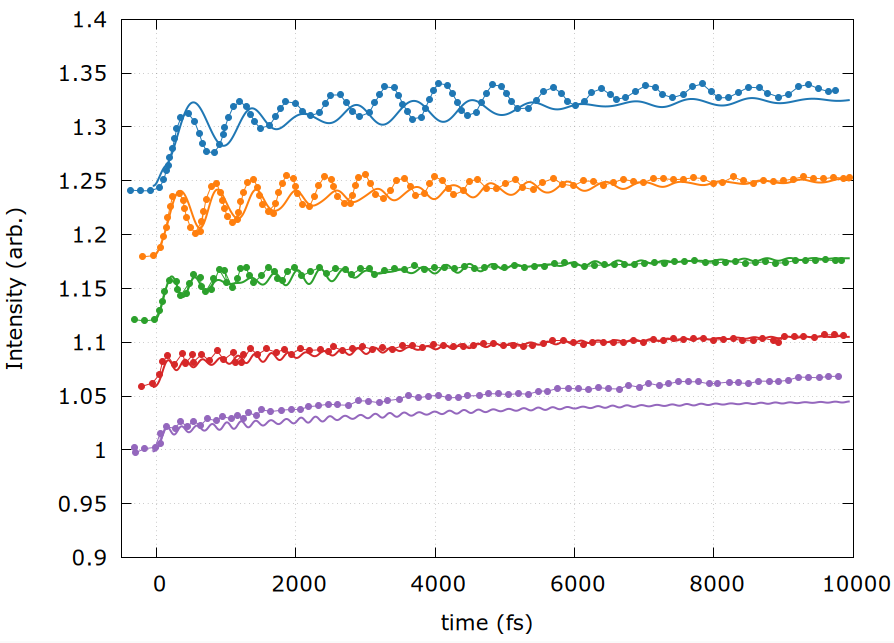}
  \caption{Calculated diffuse intensity of Fig.~\ref{fig:traces}
  ($n_\text{exc}=0.0075/\text{primitive cell}$, $T_\text{el}=\SI{1000}{\kelvin}$,
  including the phenomenological $\Gamma_\text{phen}=\SI{0.15}{\milli\electronvolt}$
  damping of Sec.~\ref{sec:numerics}) plotted directly on top of the
  experimental traces of Ref.~\onlinecite{Zhu2015}, without the vertical offset
  of Fig.~\ref{fig:traces}, so that frequency and amplitude can be compared
  trace by trace. Traces are numbered 1--5 from the smallest to the largest
  $|\mathbf{q}|$.}
  \label{fig:overlay}
\end{figure}

\begin{figure*}[htpb!]
  \centering
  \includegraphics[width=\textwidth]{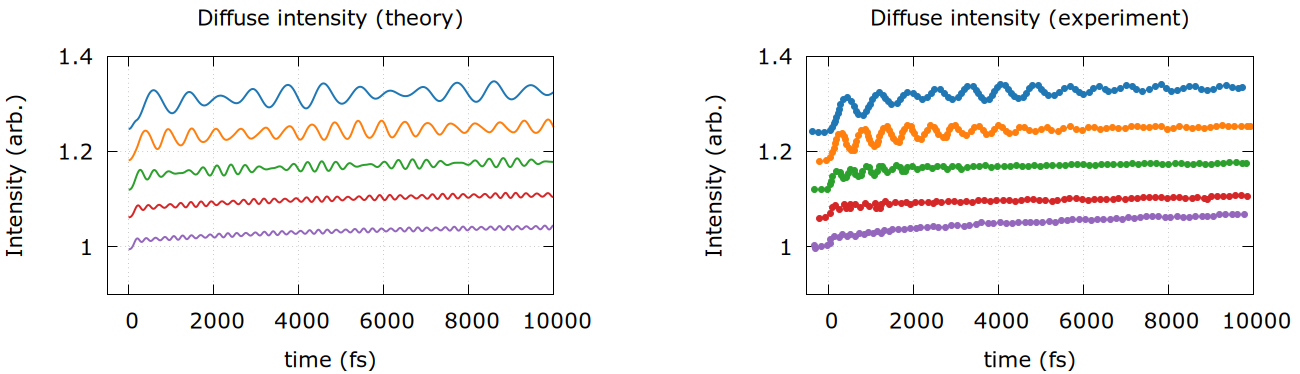}
  \caption{Calculated diffuse intensity at the five momentum transfers of
  Fig.~\ref{fig:traces}, without the additional phenomenological damping, at
  $n_\text{exc}=0.0075/\text{primitive cell}$ and
  $T_\text{el}=\SI{1000}{\kelvin}$, compared with experiment.}
  \label{fig:anharmonic}
\end{figure*}

\section{Markov approximation}
\label{app:markov}

Equations such as Eq.~\eqref{eq:app-Sdot} have the structure $\dot{y}(t) =
i\Delta(t)y(t)+f(t)$, with particular solution $y(t)=\int_{t_0}^{t}f(t')
\exp\{i\int_{t'}^{t}\Delta(t'')\,dt''\}dt'$ subject to $y(t_0)=0$, as expected
for the three-point correlators in the initial equilibrium state. Since we are dealing with
anomalous phonon correlators, the Markov approximation requires some care.

The frequency mismatch $\Delta(t)$ is determined by the natural evolution frequencies of the
operators entering the three-point correlators. Given the phonon timescales,
the initial transient over which these frequencies change appreciably can be
neglected when considering the long-time dynamics. Assuming the driving term
$f(t)$ to vary slowly compared with the oscillating phase then gives
\begin{equation}
    y(t) \simeq f(t) \left[ i\P\left(\frac{1}{\Delta(t)}\right) + \pi \delta\left(\Delta(t)\right) \right],
\label{eq:app-markov}
\end{equation}
where $\Delta(t)$ should in principle be evaluated at its quasi-equilibrium post-transient
value for the approximation to be formally valid, but is taken here to be its
instantaneous value so as to enforce energy conservation at each time during the
evolution~\cite{Stefanucci2013}.

The validity of the approximation relies on the slow variation of $f(t)$ and, in particular, of the four
phonon correlators $\langle \hat{a}_{\mathbf{q}\nu}
\hat{a}_{-\mathbf{q}\bar{\nu}}\rangle$, $\langle \hat{a}_{\mathbf{q}\nu}
\hat{a}^\dagger_{\mathbf{q}\bar{\nu}}\rangle$, $\langle
\hat{a}^\dagger_{\mathbf{q}\nu} \hat{a}_{\mathbf{q}\bar{\nu}}\rangle$, and
$\langle \hat{a}^\dagger_{\mathbf{q}\nu}
\hat{a}^\dagger_{-\mathbf{q}\bar{\nu}}\rangle$. If the quasiparticle picture
for phonons holds, their evolution due to electron-phonon scattering is slow
compared with their free dynamics, allowing the factorizations
\begin{align}
\nonumber    \langle \hat{a}_{\mathbf{q}\nu} \hat{a}_{-\mathbf{q}\bar{\nu}}\rangle &= \overline{\langle \hat{a}_{\mathbf{q}\nu} \hat{a}_{-\mathbf{q}\bar{\nu}}\rangle } \exp\left\{-i \int_{t_0}^{t}\left[\omega_{\mathbf{q}\nu}(t') + \omega_{-\mathbf{q}\bar{\nu}}(t')\right] dt' \right\}, \\
    \langle \hat{a}_{\mathbf{q}\nu} \hat{a}^\dagger_{\mathbf{q}\bar{\nu} }\rangle &= \overline{\langle \hat{a}_{\mathbf{q}\nu} \hat{a}^\dagger_{\mathbf{q}\bar{\nu} }\rangle } \exp\left\{-i \int_{t_0}^{t}\left[\omega_{\mathbf{q}\nu}(t') - \omega_{\mathbf{q}\bar{\nu}}(t')\right] dt' \right\},
\label{eq:app-factorization}
\end{align}
which separate a fast phase from a slowly evolving amplitude. For the anomalous correlator, the two phonon frequencies add in the exponential. This distinction is
irrelevant for the phonon occupations, which are already slowly varying,
but is essential to recover the correct equations of motion for phonon coherences and
other anomalous correlators within the density-matrix formalism.

Neglecting the terms involving time derivatives of the eigenmodes and
eigenfrequencies in the three-point correlator dynamics, which are relevant only during the initial transient,
and applying Eq.~\eqref{eq:app-markov} yields the Markovian expressions
for $S_{mn}^\nu$ and $R_{mn}^\nu$. Substituting these expressions back into
the equations of motion for the phonon correlators gives
Eqs.~\eqref{eq:eom-mixing-1} and~\eqref{eq:eom-mixing-2}.

\section{Comparison with experiment}
\label{app:comparison}
 
This appendix examines in greater detail the residual discrepancies between the calculation of
Sec.~\ref{sec:results} and the measurements of Ref.~\onlinecite{Zhu2015}, separating the role of electron-phonon
dephasing from that of the additional phenomenological damping and from the
choice of $T_\text{el}$ and $n_\text{exc}$.
 
\subsection{Overlay of theoretical and experimental traces}
\label{app:comparison-overlay}
 
Figure~\ref{fig:overlay} directly overlays the calculated and measured traces.
Two systematic trends emerge.

\emph{Frequency.} The oscillation period is slightly overestimated for trace~1, corresponding to the smallest $|\mathbf{q}|$ in the set.
This mode exhibits the smallest fractional photoinduced frequency shift, so the discrepancy may originate from small differences between the first-principles phonon dispersion and the experimentally measured one close to $\Gamma$. A contribution from the finite momentum
resolution $\Delta Q$ of the experiment cannot be excluded, since its relative
effect is larger for the modes closest to $\Gamma$. Nevertheless, the calculation reproduces the beating pattern observed experimentally, arising from the proximity of two phonon modes.

\emph{Background growth.} The rise of the incoherent background is
underestimated for traces~1 and~5, while traces~2--4 are reproduced more closely. Within the present
implementation, all phonon modes are constrained to follow a common time-dependent temperature,
$n_{\mathbf{q}\nu}(t)=[\exp(\hbar\omega_{\mathbf{q}\nu}(t)/k_BT(t))-1]^{-1}$,
so any genuine mode dependence of the energy-transfer rate is absent by construction. The pattern in
Fig.~\ref{fig:overlay} suggests that the experimental phonon populations develop a mode-dependent, nonthermal distribution. Describing this behaviour requires propagating the mode-resolved occupations
$n_{\mathbf{q}\nu}(t)$ beyond the common-temperature approximation, as discussed in the main text.
 
\subsection{Role of the phenomenological damping}
\label{app:comparison-anharmonic}

Section~\ref{sec:numerics} introduced an additional phenomenological rate
$\Gamma_\text{phen}$ that can be added to $\Gamma_{\mathbf{q}\nu}$ to account for
decay channels, including anharmonicity and disorder, that are not captured by the
electron-phonon damping of Eq.~\eqref{eq:eom-aa}. Figure~\ref{fig:anharmonic}
shows the first-principles result with $\Gamma_\text{phen}=0$, i.e.\ with the decay of the coherent
signal governed by the electron-phonon contribution $\Gamma_{\mathbf{q}\nu}$ alone. In the absence of the additional phenomenological damping, the oscillations persist substantially longer than in experiment, showing that electron-phonon scattering from the photoexcited carrier distribution is too weak to account for the observed coherence decay on its own. The remaining dephasing is attributed mainly to the combined effects of anharmonicity and disorder, with a possible additional contribution from the finite experimental momentum resolution.

\subsection{Sensitivity to $T_\text{el}$ and $n_\text{exc}$}
\label{app:comparison-sensitivity}

\begin{figure}[t]
  \centering
  \includegraphics[width=\columnwidth]{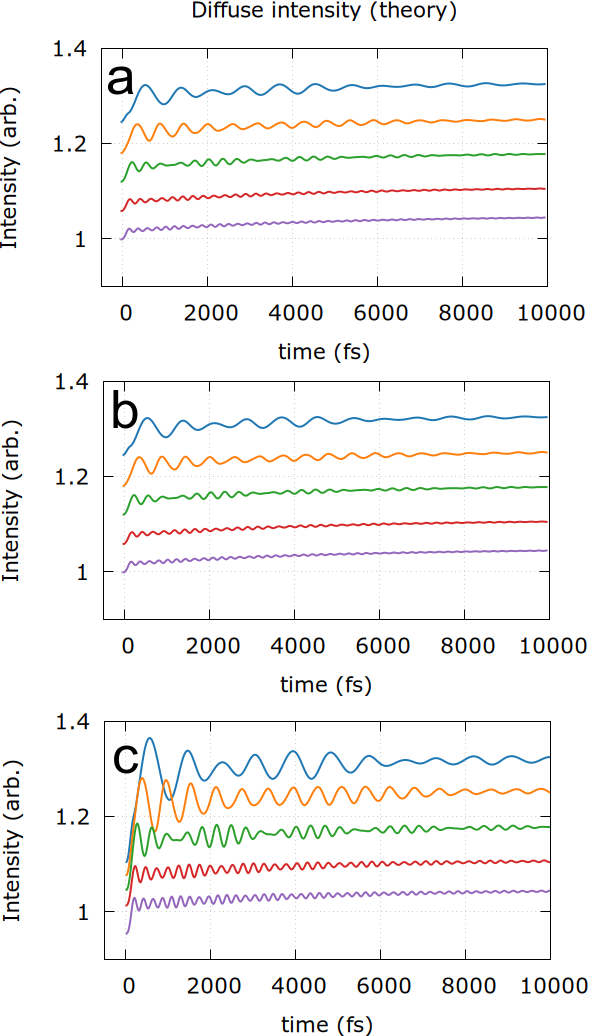}
  \caption{Calculated diffuse intensity at the five momentum transfers of
  Fig.~\ref{fig:traces} for three choices of the electronic-distribution parameters of Sec.~\ref{sec:results}: panel (a) shows the reference case of
  Fig.~\ref{fig:traces} ($T_\text{el}=\SI{1000}{\kelvin}$,
  $n_\text{exc}=0.0075/\text{primitive cell}$); panel (b) shows the same $n_\text{exc}$ with
  $T_\text{el}=\SI{2000}{\kelvin}$; and panel (c) shows $T_\text{el}=\SI{1000}{\kelvin}$ with
  $n_\text{exc}=0.025/\text{primitive cell}$.}
  \label{fig:sensitivity}
\end{figure}

We now analyze the dependence on $n_\text{exc}$ and $T_\text{el}$, which determine the
quasi-Fermi occupations entering the electron-phonon driving and damping terms. Figure~\ref{fig:sensitivity} shows the effect
of varying each parameter.

Doubling $T_\text{el}$ from \SI{1000}{\kelvin} to \SI{2000}{\kelvin} at fixed
$n_\text{exc}=0.0075/\text{primitive cell}$ leaves the oscillation frequencies essentially
unchanged. The coherent amplitude, however, increases slightly because $T_\text{el}$ modifies the electronic occupations entering the electron-phonon source and damping terms of Eq.~\eqref{eq:eom-aa}. In the present implementation, $T_\text{el}$ does not directly affect the factor $(2n_{\mathbf{q}\nu}+1)$, which is instead determined by the phonon occupations.
 
Increasing $n_\text{exc}$ from $0.0075/\text{primitive cell}$ to
$0.025/\text{primitive cell}$ at fixed $T_\text{el}=\SI{1000}{\kelvin}$ has a much stronger effect: the coherent amplitude
increases markedly because the larger excitation density enhances both the photoinduced phonon-frequency renormalization and the electron-phonon driving term. Within the sudden-quench picture, the stronger frequency renormalization is reflected in the $[1-(\omega^0/\omega^1)^2]$ factor of
Eq.~\eqref{eq:app-sudden}. The oscillation frequencies are slightly reduced, consistently with the weak photoexcitation-induced softening of the acoustic modes in this excitation-density regime.
 
\subsection{Sensitivity to the asymptotic lattice temperature}
\label{app:comparison-lattice-temperature}
 
\begin{figure}[t]
  \centering
  \includegraphics[width=\columnwidth]{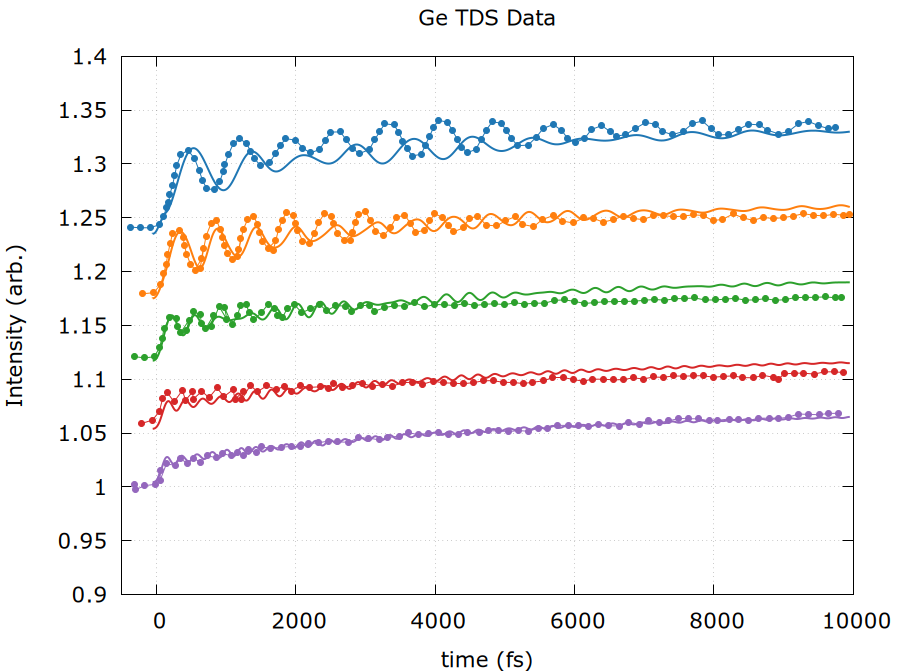}
  \caption{Calculated diffuse intensity at the five momentum transfers of
  Fig.~\ref{fig:traces} for $T_\text{latt}=\SI{315}{\kelvin}$, compared with the experimental data of Ref.~\onlinecite{Zhu2015}.}
  \label{fig:latticeT}
\end{figure}

The asymptotic lattice temperature
$T(t\to\infty)$ reached by $n_{\mathbf{q}\nu}(t)$ is a separate
parameter controlling the
long-time incoherent background through the $(2n_{\mathbf{q}\nu}+1)$ term
of Eq.~\eqref{eq:intensity}. The value used throughout this work,
$T_\text{latt}=\SI{310}{\kelvin}$, corresponds to a \SI{10}{\kelvin} increase above
the initial lattice temperature. The time traces are sensitive to this value. Figure~\ref{fig:latticeT} shows the result obtained when the asymptotic temperature
is increased to $\SI{315}{\kelvin}$, leading to a stronger rise of the calculated incoherent background. For values of $T(t\to\infty)>\SI{315}{\kelvin}$, the agreement with experiment progressively worsens (not shown).

\subsection{Fourier components of the time traces}
\label{sec:app-tds}

\begin{figure*}[t]
  \includegraphics[width=\textwidth]{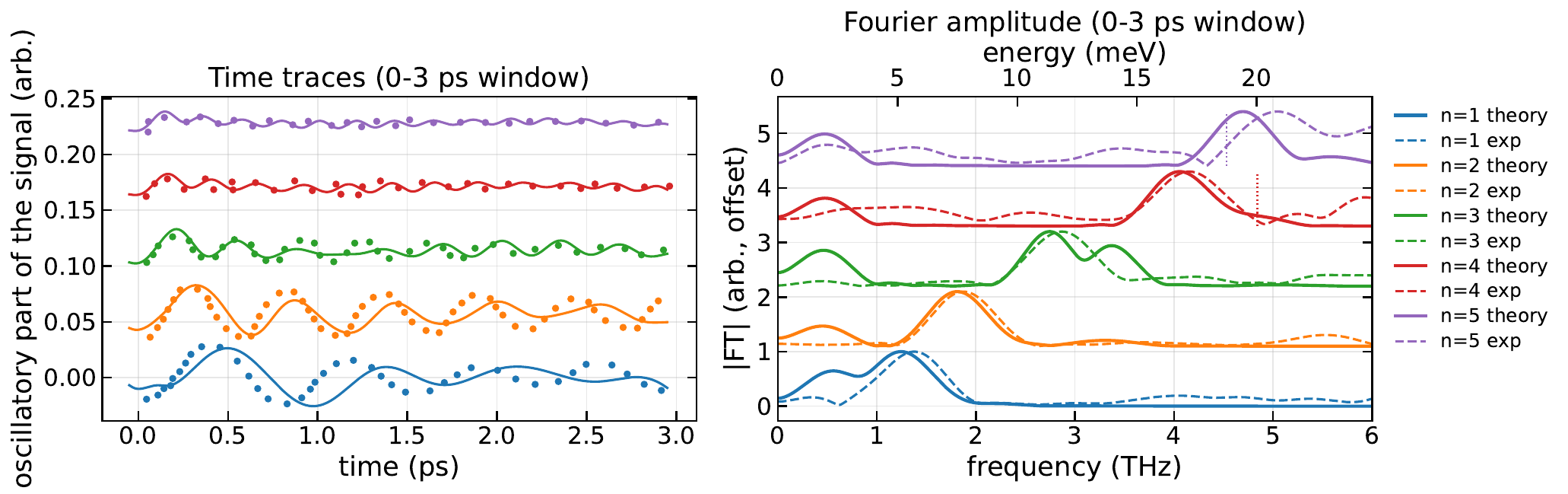}
  \caption{Fourier transforms of the oscillatory components of the curves in Fig.~\ref{fig:overlay}. The modulation occurs at twice the phonon frequency.}
  \label{fig:tds-ft}
\end{figure*}

Figure~\ref{fig:tds-ft} shows the oscillatory components of the signals in Fig.~\ref{fig:overlay} and their Fourier transforms over a time window of \SI{3}{\pico\second}.

The dominant peaks of the calculated traces reproduce the measured frequencies to within approximately $3$--$10\%$ across the accessible momentum range. A notable difference occurs for the third trace, where the calculation predicts appreciable spectral weight from both transverse-acoustic branches, whereas the experiment shows significant weight from only one. This discrepancy may partly arise from the finite momentum resolution of the experimental apparatus. The longitudinal-acoustic branch is strongly suppressed, consistently with the experimental observation that the measured signal is dominated by transverse modes.

At the two largest wavevectors, the comparison is limited by the experimental delay sampling rather than by the calculation: the corresponding modulation frequencies, $4.06$ and $4.70$~THz, lie above the Nyquist frequency associated with the experimental time sampling, indicated by the vertical dashed lines in Fig.~\ref{fig:tds-ft}.

\section{Computational details}
\label{app:computational}
 
Ground-state and photoexcited electronic structures, together with the
corresponding phonon dispersions and deformation potentials, were computed
within density-functional theory and density-functional perturbation theory as implemented in Quantum
ESPRESSO~\cite{Giannozzi2009,Giannozzi2017}, using the local-density
approximation. Norm-conserving Hartwigsen--Goedecker--Hutter (HGH)
pseudopotentials~\cite{Hartwigsen1998} were used for Ge, with
a plane-wave cutoff of \SI{80}{\ryd} for the wavefunctions. Calculations were
performed at the LDA-relaxed lattice parameter $a=\SI{5.58}{\angstrom}$, about
$1.4\%$ below the experimental value of \SI{5.658}{\angstrom} quoted in
Ref.~\onlinecite{Zhu2015}, consistent with the well-known tendency of the LDA
to overbind and hence underestimate equilibrium lattice constants.

A $12\times12\times12$ $\mathbf{k}$-mesh was used for the self-consistent and linear-response calculations. For the photoexcited calculations, a Fermi--Dirac smearing of $0.01$~Ry was used for both the electron and hole distributions, while fixed occupations were used for the ground state. An $8\times8\times8$ coarse phonon $\mathbf{q}$-mesh was used for the linear-response phonon calculations. Linear-response calculations were performed at excitation densities of $n_\text{exc}=0$, $0.025$, and $0.05$ electrons per primitive cell. Phonon frequencies at lower excitation densities were obtained by linear interpolation between the $n_\text{exc}=0$ and $n_\text{exc}=0.025$ calculations.

\begin{figure}[t]
  \centering
  \includegraphics[width=\columnwidth]{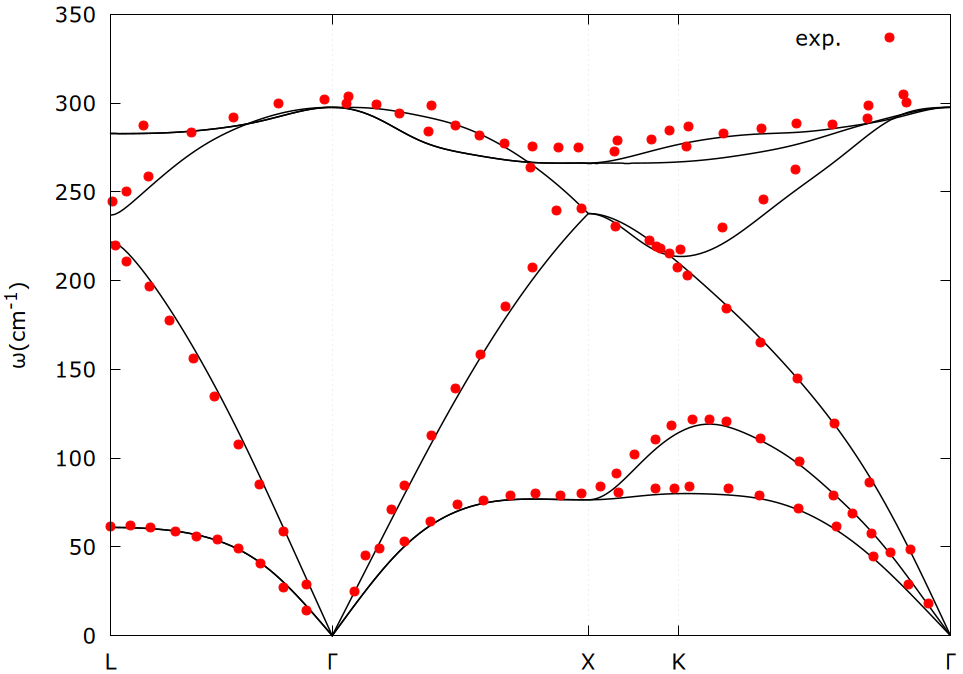}
  \caption{Calculated ground-state phonon dispersion of germanium (lines)
  along the standard high-symmetry path, compared with the inelastic neutron
  scattering data of Ref.~\onlinecite{Nilsson1971} (symbols). The comparison is used to
  validate the pseudopotential, cutoff, and lattice-parameter choices detailed
  above independently of the diffuse-scattering comparison of
  Fig.~\ref{fig:dispersion}.}
  \label{fig:groundstate-dispersion}
\end{figure}

The ground-state and photoexcited electronic structures were separately
Wannierized with Wannier90~\cite{Pizzi2020}, using $14$ Wannier functions per
formula unit spanning the valence manifold together with the eight lowest
conduction states relevant to the \SI{1.55}{\electronvolt} pump photon energy
of Ref.~\onlinecite{Zhu2015}. The resulting Wannier Hamiltonians and
deformation-potential matrix elements were Fourier-interpolated onto
dense $\mathbf{k}$- and $\mathbf{q}$-grids within \textsc{EPI}q, as
summarized in Fig.~\ref{fig:workflow}.

The photoexcited Born--Oppenheimer surface, characterized by $\omega^1_{\mathbf{q}\nu}$, together with the
corresponding deformation potentials of Sec.~\ref{sec:implementation}, was
obtained self-consistently at fixed excitation density $n_\text{exc}$ using
the two-chemical-potential extension of the Kohn--Sham
self-consistency cycle implemented in Quantum
ESPRESSO~\cite{PhysRevB.75.184301,Marini2021}. Separate quasi-Fermi levels were assigned
to the valence and conduction manifolds and adjusted self-consistently to
reproduce the prescribed $n_\text{exc}$ at a given electronic temperature.

Figure~\ref{fig:groundstate-dispersion} compares the calculated ground-state phonon dispersion with the measured neutron-scattering dispersion of Ge at \SI{80}{\kelvin}. The agreement is good across the
acoustic branches relevant to this work.

The time-dependent simulations were performed using $30\times30\times30$ dense $\mathbf{k}$- and $\mathbf{q}$-grids and a smearing of $1$~mRy for the energy-conserving delta functions. The momentum-transfer points were chosen according to Fig.~2 of Ref.~\onlinecite{Zhu2015}, within the Brillouin zone centred on the $(022)$ reciprocal-lattice vector. Photoexcited electron-phonon matrix elements obtained within the two-chemical-potential approach were used throughout the calculations, thereby accounting for the photoinduced modification of the electron-phonon coupling in the presence of free carriers. The theoretical time traces were shifted by $-50$~fs relative to the nominal experimental time zero to account phenomenologically for the initial pump-excitation stage.

\bibliographystyle{apsrev4-2}
\bibliography{refs}

\end{document}